# Optimization Algorithms for Faster Computational Geometry

Zeyuan Allen-Zhu
zeyuan@csail.mit.edu
Princeton University

Zhenyu Liao
zhenyul@bu.edu
Boston University

Yang Yuan
yangyuan@cs.cornell.edu
Cornell University

**Abstract**

We study two fundamental problems in computational geometry: finding the maximum inscribed ball (MaxIB) inside a bounded polyhedron defined by $m$ hyperplanes, and the minimum enclosing ball (MinEB) of a set of $n$ points, both in $d$-dimensional space. We improve the running time of iterative algorithms on

MaxIB from $\widetilde{O}(md\alpha^3/\varepsilon^3)$ to $\widetilde{O}(md + m\sqrt{d}\alpha/\varepsilon)$, a speed-up up to $\widetilde{O}(\sqrt{d}\alpha^2/\varepsilon^2)$, and[1]

MinEB from $\widetilde{O}(nd/\sqrt{\varepsilon})$ to $\widetilde{O}(nd + n\sqrt{d}/\sqrt{\varepsilon})$, a speed-up up to $\widetilde{O}(\sqrt{d})$ .

Our improvements are based on a novel saddle-point optimization framework. We propose a new algorithm `L1L2SPSolver` for solving a class of regularized saddle-point problems, and apply a randomized Hadamard space rotation which is a technique borrowed from compressive sensing. Interestingly, the motivation of using Hadamard rotation solely comes from our optimization view but not the original geometry problem: indeed, it is not immediately clear why MaxIB or MinEB, as a geometric problem, should be easier to solve if we rotate the space by a unitary matrix. We hope that our optimization perspective sheds lights on solving other geometric problems as well.

## 1 Introduction

The goal of this paper is to bridge the fields of optimization and computational geometry using a simple unified saddle-point framework. As two immediate products of this new connection, we obtain faster iterative algorithms to approximately solve two fundamental problems in computational geometry: the maximum inscribed ball problem (MaxIB) and the minimum enclosing ball problem (MinEB). Our methods are composed of simple updating rules on vectors and therefore do not require geometric operations that are found in classical algorithms. This is another example of surprisingly good results obtained using optimization insights following the current trend of theoretical computer science.

In the rest of this introduction, we describe the definitions of the MaxIB and MinEB problems and review prior work. In the next three sections, we describe our saddle-point formulation and algorithms for MaxIB and MinEB.

---

[*]The first version of this paper appeared in December 2014 but contains only the smooth convex optimization based algorithms. The second version of this paper appeared in December 2015 and already contains all the technical details of this present paper.

[1]$\alpha \geq 1$ is the aspect ratio of the polyhedron. Throughout this paper we use the $\widetilde{O}$ notation to hide logarithm factors such as $\log m, \log d, \log \alpha$, and $\log(1/\varepsilon)$.

**Maximum Inscribed Ball (MaxIB).** In the MaxIB problem, we are given a polyhedron $P$ in $\mathbb{R}^d$ defined by $m$ halfspaces $\{H_1, \ldots, H_m\}$. Each halfspace $H_j$ is characterized by a linear constraint $\langle A_j, x \rangle + b_j \geq 0$. As in prior work [XSX06], we assume that $P$ is bounded (so $m \geq d$) and a common point is known to be contained in $P$ – without loss of generality, let it be the origin $O$. Let $\alpha \geq 1$ be an upper bound on the aspect ratio of $P$, i.e., the ratio between the radii of the minimum enclosing ball and the maximum inscribed ball of $P$, and $\varepsilon > 0$ be a desired error bound.

The goal of MaxIB is to find a point $x \in P$ such that its minimum distance to all the bounding hyperplanes $H_j$ is at least $(1-\varepsilon)r_{\mathsf{opt}}$, where $r_{\mathsf{opt}}$ is the radius of a maximum inscribed ball of $P$.

Besides the applications in computational geometry, MaxIB has also been used in the column generation method [LP11] and the sphere method [Mur12] for linear programming, and the central cutting-plane method for convex programming [EM75].

When the dimension is a constant, the $\varepsilon$-kernel technique (see the survey [AHV05]) yields a linear-time approximation algorithm for MaxIB based on core-set construction. However, its running time is proportional to $\varepsilon^{-\Omega(d)}$. In high dimensions, finding the maximum inscribed ball remains a challenging problem in theoretical computer science and operations research. One can reduce this problem to a linear program [EM75] and rely on existing LP solvers, however, the so-obtained algorithm can be too slow for practical purposes (although still in polynomial time).

In an influential paper, Xie, Snoeyink, and Xu [XSX06] obtained an approximation algorithm for MaxIB with running time $O(md\alpha^3/\varepsilon^3 + md\alpha \log \alpha) = \widetilde{O}(md\alpha^3/\varepsilon^3)$. Their algorithm is based on a number of interesting geometric observations, as well as a dual transformation to reduce the MaxIB problem to a sequence of minimum enclosing ball (MinEB) instances, which they solve by applying known core-set techniques [BHI02, KMY03]. Unfortunately, their cubic dependence on $\alpha$ and $1/\varepsilon$ undermines the practical applicability of their algorithm.

In Section 3, we use saddle-point optimization techniques to obtain an algorithm `MaxIBSPSolver` with running time $\widetilde{O}(md + m\sqrt{d}\alpha/\varepsilon)$. In other words, we reduce the dependence on both $\alpha$ and $1/\varepsilon$ from cubic to linear, and improve the running time by a factor up to $\sqrt{d}\alpha^2/\varepsilon^2$. We emphasize that our improvement could be significant in the views of theoretical computer scientists, operations researchers, as well as experimentalists:

- In theoretical computer science, one usually views $\alpha$ and $\varepsilon$ as large constants so our improvement can be seen as $\widetilde{\Omega}(\sqrt{d})$ if one ignores the input reading time $O(md)$.

- In operations research or statistics, one usually concentrate on the convergence rate which is the $\varepsilon$ dependence (recall that the seminal work of Nesterov is only to reduce $1/\varepsilon$ to $1/\sqrt{\varepsilon}$ [Nes83]). Our improvement in this paper is from $1/\varepsilon^3$ to $1/\varepsilon$.

- In practice, if $\alpha$ is 10 for the polyhedron, $\varepsilon$ is 10%, and the dimension $d = 100$, our method could potentially be $10^5$ times faster than that of [XSX06]. We leave it a future work to inspect the practical performance of our method on real-life datasets.

In Appendix B, we also apply convex (rather than saddle-point) optimization and obtain a *parallel* algorithm `MaxIBConvexSolver` with slightly slower *total* running time $\widetilde{O}(md\alpha/\varepsilon)$. However, in terms of *parallel* running time (i.e., the number of parallelizable iterations, a classical benchmark used by iterative solvers [AO15b]), `MaxIBConvexSolver` improves the result of [XSX06] by a factor $\widetilde{\Omega}(\alpha^2/\varepsilon^2)$.

**Minimum Enclosing Ball (MinEB).** In the MinEB problem, we are given a set $\{a_1, a_2, \ldots, a_n\} \subseteq \mathbb{R}^d$ of points in the $d$-dimensional space and are asked to find a point $x \in \mathbb{R}^d$ so that its maximum distance to all the $n$ points is at least $(1+\varepsilon)R_{\mathsf{opt}}$, where $R_{\mathsf{opt}}$ is the radius of a minimum enclosing ball that contains all the points in this set.

As originally studied by Sylvester in [Syl57], the problem of MinEB has found numerous applications in fields such as data mining, learning, statistics, and computer graphics. In particular, the relationship between MinEB and support vector machines (SVMs) has been recently emphasized by [HRZ07, GJ09, Cla10, SVZ11]. Efficient algorithms for this problem are both of theoretical and practical importance.

If the dimension $d$ is constant, the algorithm of Welzl [Wel91] solves MinEB exactly in linear time. Unfortunately, its dependency on $d$ is exponential.

For large dimensions, a sequence of works based on the core-set technique [BHI02, KMY03, BC08, Yil08, Cla10] has given algorithms whose best known running time is $O(nd/\varepsilon)$. This running time is tight for the core-set technique, as, in the worst-case, the size of a coreset of MinEB is at least $\Omega(1/\varepsilon)$ [BC08]. Another type of algorithm due to Clarkson, Hazan, and Woodruff [CHW12] achieves a running time of $\widetilde{O}(n/\varepsilon^2 + d/\varepsilon)$. This algorithm is fast for large values of $\varepsilon$, but may not be suitable for very small $\varepsilon$. All these cited algorithms converge at best in $O(1/\varepsilon)$ iterations.

Recently, Saha, Vishwanathan, and Zhang [SVZ11] designed two algorithms for MinEB that successfully overcame this $1/\varepsilon$ barrier. Using our $\varepsilon$-notation for multiplicative error, they give one algorithm which works in the $\ell_2$-norm and achieves a running time of $O(ndQ/\sqrt{\varepsilon})$, and another algorithm which works in the $\ell_1$-norm and achieves a running time of $O(nd\sqrt{\log n}L/\sqrt{\varepsilon})$. While the values of $Q$ and $L$ depend on the input structure, we observe that $Q$ can be as large as $O(\sqrt{n})$, while $L$ is never larger than a constant. In other words, their proposed algorithms have worst-case running times $O(n^{1.5}d/\sqrt{\varepsilon})$ and $O(nd\sqrt{\log n}/\sqrt{\varepsilon})$. The key component behind the result of Saha, Vishwanathan, and Zhang is the excessive gap framework of Nesterov [Nes05a], which is a primal-dual first-order approach for structured non-smooth optimization problems.

In Section 4, we rewrite MinEB as a saddle-point optimization problem, and obtain an algorithm `MinEBSPSolver` that runs in $\widetilde{O}(nd + n\sqrt{d}/\sqrt{\varepsilon})$. This is faster than the previous algorithm [SVZ11] by a factor up to $\sqrt{d}$, and faster than the popular core-set algorithm by a factor up to $\sqrt{d}/\sqrt{\varepsilon}$.

As an additional result, in Appendix D, we also observe that MinEB can be directly formulated as a convex (rather than saddle-point) optimization problem, and get an algorithm `MinEBConvexSolver` matching the running time of [SVZ11] but with much simpler analysis.

**Remark.** For both MaxIB and MinEB, one can also use interior-point types of algorithms to obtain a convergence rate of $\log(1/\varepsilon)$. However, this fast convergence rate comes at the cost of having expensive iterations: each iteration typically requires solving a linear equation system in the input size, making it impractical for very-large-scale inputs. Therefore, in this paper, we choose to focus on iterative methods whose iterations run in nearly-linear time.

## 1.1 Our Techniques

Our `MaxIBSPSolver` and `MinEBSPSolver` rely on (min-max) saddle-point optimization to solve MaxIB and MinEB respectively. More specifically, we reduce MaxIB and MinEB to solving the regularized saddle-point program:

$$\max_{x \in \mathbb{R}^d} \min_{y \in \Delta_m} \frac{1}{d} y^T A x + \frac{1}{d} y^T b + \lambda H(y) - \frac{\gamma}{2} \|x\|_2^2 \ ,$$

where $H(\cdot)$ is the entropy function over $m$-dimensional probabilities vectors, and $\lambda, \gamma > 0$ are fixed regularization parameters. We call this $\ell_1$-$\ell_2$ saddle-point optimization because, borrowing language from optimization, this objective is strongly convex with respect to the $\ell_1$ norm on the $y$ side and strongly concave with respect to the $\ell_2$ norm on the $x$ side.

To solve this saddle-point problem efficiently, we iteratively update $x$ and $y$. In particular, in each iteration we update $x$ by a *random coordinate*, and update $y$ fully using multiplicative weight updates. Therefore, this method can be viewed as an *accelerated, coordinate-based*, first-order method for saddle-point optimization. To the best of our knowledge, the only previously known accelerated, coordinate-based method on saddle-point optimization was SPDC [ZX15], one of the state-of-the-art algorithms used for empirical risk minimizations in machine learning. We call our algorithm `L1L2SPSolver`.

**A Surprising Hadamard Rotation.** Unfortunately, solely applying `L1L2SPSolver` does not solve MinEB or MaxIB fast enough. In particular, the running time of `L1L2SPSolver` relies on the largest absolute values of $A$'s entries. If the entries of $A$ are very non-uniform —say, with a few very large entries and mostly small ones— the performance could be somewhat unsatisfactory. (In particular, we no longer have a $\sqrt{d}$ factor speed-up.)

To overcome this difficulty, we apply a randomized Hadamard transformation (Lemma 2.1) on $A$ to uniformize its entries, so that all entries of $A$ are relatively small. This transformation is inspired by the fast Johnson-Lindenstrauss transform [AC10] proposed for numerical linear algebra and compressive sensing purposes, and is another crucial ingredient behind our running time improvements.

Surprisingly, this Hadamard rotation comes solely from our optimization view but not the geometry. Indeed, it is not immediately clear why MaxIB or MinEB, as geometric problems, should be easier to solve if we rotate the space by a unitary (Hadamard) matrix.

**Our Contributions.** We summarize the main contributions of this paper as follows:

- We provide significantly faster algorithms on MaxIB and MinEB.

- This is the first time coordinate-based saddle-point optimization algorithm is applied to MaxIB, MinEB, or perhaps to any computational geometry problem.

- Since the $\ell_1$-$\ell_2$ saddle-point problem seems very natural, our `L1L2SPSolver` method can potentially lead to other applications in the future.

- The speed-up we obtained from the Hadamard rotation is an algebraic technique but applied to geometric problems. It sheds lights on solving perhaps more geometric problems faster using optimization insights.

## 2 $\ell_1$-$\ell_2$ Saddle Point Optimization

In this section we study the following saddle-point problem that may be of independent interest. We shall later use it to solve MaxIB and MinEB.

$$\max_{x \in \mathbb{R}^d} \min_{y \in \Delta_m} \frac{1}{d} y^T A x + \frac{1}{d} y^T b + \lambda H(y) - \frac{\gamma}{2} \|x\|_2^2 \ . \tag{2.1}$$

Above, $A \in \mathbb{R}^{m \times d}$ is a given matrix, $b \in \mathbb{R}^m$ is a given vector, $\lambda, \gamma > 0$ are two regularization parameters, and the entropy function $H(y) \stackrel{\text{def}}{=} \sum_{i=1}^m y_i \log y_i$ is defined over $\Delta_m$, the set of $m$-dimensional probabilities vectors.

Define the Bregman divergence function $V_x(y) \stackrel{\text{def}}{=} H(y) - \langle \nabla H(x), y - x \rangle - H(x) = \sum_i y_i \log(\frac{y_i}{x_i})$. It is a known fact that $H(\cdot)$ is 1 strongly convex with respect to the $\ell_1$ norm, or in symbols, for every $x, y \in \Delta_m$, we have $V_x(y) \geq \frac{1}{2}\|x - y\|_1^2$. Therefore, our objective (2.1) is $\lambda$ strongly convex with respect to the $\ell_1$ norm on the $y$ side, and $\gamma$ strongly concave with respect to the $\ell_2$ norm on the $x$ side.

**Algorithm 1** `L1L2SPSolver`$(A, b, \lambda, \gamma, T)$

**Input:** Matrix $A \in \mathbb{R}^{m \times d}$, vector $b \in \mathbb{R}^m$, parameters $\lambda, \gamma > 0$, number of iterations $T$.
   Assume that $\sum_{j=1}^{d} A_{ij}^2 \leq 1$ for each $i \in [m]$.
**Output:** $x^{(T)}, y^{(T)}$ that approximately solve (2.1).
1: Assume without loss of generality that $|A_{j,i}| \leq q/\sqrt{d}$ for some parameter $q = O(\sqrt{\log m})$.
   $\diamond$ *See Lemma 2.1*
2: $\tau \leftarrow \frac{1}{2q}\sqrt{\frac{d\gamma}{\lambda}}$, $\sigma \leftarrow \frac{1}{2q}\sqrt{\frac{d\lambda}{\gamma}}$, and $\theta \leftarrow 1 - \frac{1}{d+q/\sqrt{\lambda d\gamma}}$.
3: $x^{(0)} \leftarrow (0, \cdots, 0)$, and $y^{(-1)} = y^{(0)} \leftarrow (\frac{1}{m}, \cdots, \frac{1}{m})$.
4: **for** $t \leftarrow 0$ **to** $T - 1$ **do**
5:    Pick an index $i^* \in \{1, 2, \cdots, d\}$ uniformly at random
6:    $\forall i \in [d], x_i^{(t+1)} \leftarrow \begin{cases} (\sigma d\gamma + 1)^{-1}(x_i^{(t)} + \sigma \langle y^{(t)} + \theta(y^{(t)} - y^{(t-1)}), A_i \rangle) & \text{if } i = i^*, \\ x_i^{(t)} & \text{if } i \neq i^*. \end{cases}$
7:    $\forall j \in [m], y_j^{(t+1)} \leftarrow \frac{1}{Z} \exp \left\{ \frac{\frac{1}{\tau} \log y_j^{(t)} - \frac{1}{d}\left(A(x^{(t)} + d(x^{(t+1)} - x^{(t)}))\right)_j - \frac{b_j}{d}}{\lambda + \frac{1}{\tau}} \right\}$
   $\diamond$ *where $Z > 0$ is the normalizer that ensures $\sum_{j \in [m]} y_j^{(t+1)} = 1$*
8: **end for**
9: **return** $x^{(T)}, y^{(T)}$.

---

We denote by $x^\circ, y^\circ$ the optimal saddle point of this objective (2.1). In this section, we view $A \in \mathbb{R}^{m \times d}$ as $A = [A_1, \ldots, A_d]$ where each $A_i$ is an $m$-dimensional vector. We assume without loss of generality that $\sum_{i=1}^{d} A_{ji}^2 \leq 1$ for each $j \in [m]$.

The above assumption only implies $|A_{ji}| \leq 1$ for all pairs of $j, i$. However, if we rotate the space randomly, we can assume $|A_{ji}| \leq O(\sqrt{\log m/d})$ without loss of generality:

**Lemma 2.1** (Hadamard Transform). *There exist $q = O(\sqrt{\log m})$ and a unitary matrix (i.e., a $d$-dimensional rotation) $T \in \mathbb{R}^{d \times d}$, such that $|(AT)_{ji}| \leq q/\sqrt{d}$ for all $i \in [d]$ and $j \in [m]$. Setting $A' \stackrel{\text{def}}{=} AT$ and $x' \stackrel{\text{def}}{=} T^{-1}x$, this reduces (2.1) to a new saddle-point optimization problem*

$$\max_{x' \in \mathbb{R}^d} \min_{y \in \Delta_m} \frac{1}{d} y^T A' x' + \frac{1}{d} y^T b + \lambda H(y) - \frac{\gamma}{2} \|x'\|_2^2$$

*without changing the solution (up to the unitary transformation). Moreover, $TA$ can be computed in expected running time $O(md \log d)$.*

*Proof.* Assume without loss of generality that $d$ is a power of 2 — otherwise one can certainly add a few dummy dimensions without changing the solution of the problem. Next, let $H$ be the $d \times d$ Walsh-Hadamard matrix, and $D$ be a $d \times d$ diagonal matrix whose entries are i.i.d. chosen from $\pm 1$. It is a well celebrated result in compressed sensing that with probability at least, say, 0.95, it satisfies that (see for instance [AC10, Equation (3)])

$$\forall j \in [m], i \in [d], \quad |(HDA^T)_{ij}| \leq O(\sqrt{\log m/d}) \ . \tag{2.2}$$

Since $HD$ is a $d \times d$ unitary transformation of the space, we can define $T = (HD)^T$ and this provides the desired reduction. The running time needed to compute $A(HD)^T$ is only $O(md \log d)$ using FFT (see for instance [AC10] again). Although this reduction succeeds with probability only 0.95, if we fail, we can re-generate $D$ until $|A'_{ji}| \leq O(\sqrt{\log m/d})$ for all pairs $j, i$. $\square$

Owing to the above lemma we simply assume that a rotation is already applied so $A$ satisfies $|A_{ji}| \leq O(\sqrt{\log m/d})$ in this paper. We propose `L1L2SPSolver`$(A, b, \lambda, \gamma, T)$ in Algorithm 1 to solve (2.1). We have the following theorem:

> **Theorem 2.2.** *Assuming $\sum_{j=1}^{d} A_{ij}^2 \leq 1$ for each $i \in [m]$, the outputs $x^{(T)}, y^{(T)} = $ `L1L2SPSolver`$(A, b, \lambda, \gamma, T)$ satisfy*
>
> $$\left(\frac{1}{\tau} + 2\lambda\right) \mathbb{E}\left[V_{y^{(T)}}(y^\circ)\right] + \left(\frac{1}{4\sigma} + d\gamma\right) \mathbb{E}[\|x^\circ - x^{(T)}\|_2^2]$$
> $$\leq \theta^T \cdot \left(\left(\frac{1}{\tau} + 2\lambda\right) \log m + \left(\frac{1}{2\sigma} + d\gamma\right) \|x^\circ\|_2^2\right), \quad (2.3)$$
>
> *where parameters $\tau = \frac{1}{2q}\sqrt{\frac{d\gamma}{\lambda}}$, $\sigma = \frac{1}{2q}\sqrt{\frac{d\lambda}{\gamma}}$, $\theta = 1 - \frac{1}{d + q/\sqrt{\lambda d\gamma}}$ for some $q = O(\sqrt{\log m})$.*

Theorem 2.2 can be interpreted as follows: $x^{(T)}, y^{(T)}$ converge to their corresponding optimums $x^\circ, y^\circ$ with an exponential rate $\theta^T$. Indeed, if $\|x^\circ - x^{(T)}\|_2$ approaches zero then $x^{(T)}$ approaches $x^\circ$; similarly if $V_{y^{(T)}}(y^\circ)$ approaches zero then $y^{(T)}$ approaches $y^\circ$.

The proof of Theorem 2.2 is very technical and deferred to Appendix A.

We remark here that although the algorithm looks very different from known methods in the saddle-point optimization literatures (such as [CP11, Nem04, ZX15]), readers with rich optimization background may notice that `L1L2SPSolver` is a generalization of SPDC [ZX15], a stochastic saddle-point optimization method that has recently become popular in the machine learning community.

At a high level, SPDC only supports Euclidean norms, where our objective (2.1) is strongly convex with respect to the $\ell_1$ norm on the $y$ side. To properly deal with $\ell_1$ norms, we perform exponential updates on the $y$ side, which can be viewed as *multiplicative weight updates* which have been recently found very useful for theoretical computer science applications (see for instance [AHK12] for a survey and [AO14, AO15a, AO15b, AZLO16] for connections to optimization).

## 3 MaxIB: From Geometry to Saddle-Point Optimization

In this section, we study the problem of finding the maximum inscribed ball (MaxIB) inside a bounded polyhedron defined by $m$ hyperplanes in a $d$-dimensional space, using `L1L2SPSolver` proposed in Section 2. Below we first write MaxIB as an optimization problem, and prove some basic facts in Section 3.1, then apply `L1L2SPSolver` in Section 3.2.

### 3.1 Optimization View of MaxIB

Without loss of generality, we assume $\sum_{i=1}^{d} A_{ji}^2 = 1$ for each $j \in [m]$. We shall view $A$ as an $m \times d$ matrix whose rows consist of $A_j$ for $j = 1, \ldots, m$. Throughout this paper, we interchangeably view $H_j$ both as a halfspace and as a hyperplane. The *directed* distance from any point $x$ to the separating hyperplane $H_j$ is $\langle A_j, x \rangle + b_j$: when this value is negative, it indicates that $x$ is outside the halfspace $H_j$, and positive vice versa. In particular, the distance from the origin $O$ to hyperplane $H_j$ is $b_j \geq 0$, because the origin $O$ is assumed to be inside the polyhedron $P$.

Let $B \stackrel{\text{def}}{=} \max_i \{b_i\}$, and denote by $r_{\mathsf{opt}}$ the radius of the maximum inscribed ball of $P$, by $R$ the radius of the minimum enclosing ball (MinEB) of $P$, and by $x^*$ the center of (any) maximum inscribed ball. By the definition of aspect ratio, we have $R \leq \alpha \cdot r_{\mathsf{opt}}$. We first note a simple fact from geometry, which is a consequence of the boundedness of the polyhedron:

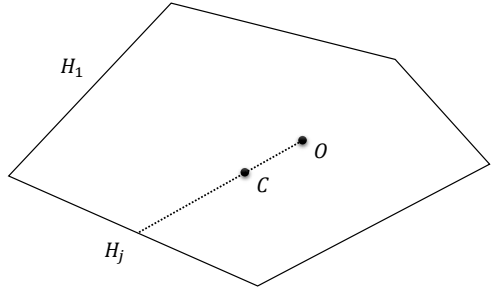

Figure 1: Illustration for the proof of Fact 3.1.

**Fact 3.1.** $r_{\mathsf{opt}} \leq B$.

*Proof.* As shown in Figure 1, let $O$ be the origin and $C$ be the center of any maximum inscribed ball of $P$. Let us now connect $\overrightarrow{OC}$ and prolong it until the line hits some hyperplane $H_j$. Next, we compute that $B \geq b_j = \mathrm{dist}(O, H_j) \geq \mathrm{dist}(C, H_j) \geq r_{\mathsf{opt}}$, concluding the proof of Fact 3.1. □

Let $\Delta_m$ denote the unit simplex in $\mathbb{R}^m$, that is, $\Delta_m = \{x \in \mathbb{R}^m \colon x \geq 0 \wedge \mathbb{1}^T x = 1\}$. Then, we can write our computational geometry problem into the following saddle-point problem:

**Lemma 3.2.** $r_{\mathsf{opt}} = \max_{x \in \mathbb{R}^d} \min_{y \in \Delta_m} y^T(Ax + b)$.

*Proof.* Since for any point $x \in \mathbb{R}^d$ the directed distance from $x$ to a hyperplane $H_j$ is $\langle A_j, x \rangle + b_j$, the minimum $\min_{j \in [m]}\{\langle A_j, x \rangle + b_j\}$ is equal to the maximum radius of a ball centered at $x$ that is contained in $P$. (Or, if this value is negative, it means $x$ is outside $P$.) Therefore, for our MaxIB problem it suffices to maximize $\min_{j \in [m]}\{\langle A_j, x \rangle + b_j\}$ over all possible $x \in \mathbb{R}^d$. That is,

$$r_{\mathsf{opt}} = \max_{x \in \mathbb{R}^d} \min_{j \in [m]} \langle A_j, x \rangle + b_j = \max_{x \in \mathbb{R}^d} \min_{y \in \Delta_m} y^T(Ax + b) \ .$$

Above, the second equality holds because $\min_{j \in [m]} \langle A_j, x \rangle + b_j = \min_{y \in \Delta_m} y^T(Ax + b)$. □

We also have the following handy sandwich lemma:

**Lemma 3.3.** $\|x^*\| \leq 2\alpha r_{\mathsf{opt}} \leq 2\alpha\beta$.

*Proof.* Since the center $x^*$ of MaxIB is in $P$, the segment from $O$ to $x^*$ lies completely inside $P$ and thus also inside the minimum enclosing ball of $P$. Therefore, we have $\|x^*\| \leq 2R \leq 2\alpha r_{\mathsf{opt}} \leq 2\alpha\beta$. □

## 3.2 Apply `L1L2SPSolver` to MaxIB

In Lemma 3.2 we have already characterized the MaxIB problem as a saddle point optimization $r_{\mathsf{opt}} = \max_{x \in \mathbb{R}^d} \min_{y \in \Delta_m} y^T(Ax + b)$. Accordingly, define

$$\phi(x, y) \stackrel{\mathrm{def}}{=} y^T A x + y^T b \quad \text{and} \quad f(x) \stackrel{\mathrm{def}}{=} \min_{y \in \Delta_m} \phi(x, y) \ ,$$

and let $x^*, y^*$ be an optimal saddle point of this above saddle point problem.[2] Now, it suffices for us to find a point $x$ such that $f(x) \geq (1-\varepsilon) r_{\mathsf{opt}} = (1-\varepsilon) f(x^*)$ because such an $x$ is necessarily a $(1-\varepsilon)$-approximate solution to MaxIB.

---
[2]Note that $x^*$ is also necessarily a maximizer of $f(x)$.

Unfortunately, $\phi(x,y)$ does not fall into the category of (2.1) since it is not strongly convex (resp. concave) with respect to $y$ (resp. $x$). For this reason, we define the following regularized saddle-point problem

$$\max_{x\in\mathbb{R}^d}\min_{y\in\Delta_m}\left\{\phi_r(x,y)\stackrel{\text{def}}{=} y^T Ax + y^T b + d\lambda H(y) - \frac{d\gamma}{2}\|x\|_2^2\right\},\tag{3.1}$$

where the parameters

$$\lambda \stackrel{\text{def}}{=} \frac{\epsilon(\beta/c)}{4d\cdot\log m} \quad\text{and}\quad \gamma \stackrel{\text{def}}{=} \frac{\epsilon}{2d\cdot 4\alpha^2\beta}\ .$$

Above, recall that $\alpha > 0$ is a known (upper bound) on the aspect ratio of the polyhedron, and $\beta$ is some constant approximation of $r_{\text{opt}}$ that can be obtained from preprocessing and satisfies $\beta/c \leq r_{\text{opt}} \leq \beta$ for constant $c > 1$.[3]

We now make a few claims before we apply our saddle-point algorithm in Section 2.

**Claim 3.4.** $\lambda \leq \frac{\epsilon r_{\text{opt}}}{4d\cdot\log m} \quad\text{and}\quad \gamma \leq \frac{\epsilon r_{\text{opt}}}{2d\cdot\|x^*\|_2^2}\ .$

*Proof.* The first inequality is a direct consequence of $\beta/c \leq r_{\text{opt}}$. To prove the second inequality, recall Lemma 3.3 shows $\|x^*\| \leq 2\alpha r_{\text{opt}} \leq 2\alpha\beta$. Therefore, we have $1/(4\alpha^2\beta) \leq \beta/\|x^*\|^2 \leq r_{\text{opt}}/\|x^*\|^2$. □

Denote by $(x^\circ, y^\circ)$ the optimal saddle-point of objective (3.1). Then, we have

**Claim 3.5.** $f(x^*) - f(x^\circ) \leq \frac{\epsilon r_{\text{opt}}}{2}\ .$

*Proof.* Denoting $\widetilde{y} \stackrel{\text{def}}{=} \arg\min_{y\in\Delta_m}\phi(x^\circ, y)$, we have (by the definition of saddle points),

$$f(x^\circ) = \widetilde{y}^T Ax^\circ + \widetilde{y}^T x^\circ \geq \phi_r(x^\circ, \widetilde{y}) - d\lambda H(\widetilde{y}) \geq \phi_r(x^\circ, y^\circ) - d\lambda H(\widetilde{y})$$

$$\geq \phi_r(x^*, y^\circ) - d\lambda H(\widetilde{y}) \geq \phi(x^*, y^\circ) - d\lambda H(\widetilde{y}) - \frac{d\gamma}{2}\|x^*\|_2^2$$

$$\geq \phi(x^*, y^*) - d\lambda H(\widetilde{y}) - \frac{d\gamma}{2}\|x^*\|_2^2 = f(x^*) - d\lambda H(\widetilde{y}) - \frac{d\gamma}{2}\|x^*\|_2^2\ .$$

Since $H(\widetilde{y}) \in [0, \log m]$ we immediately get $f(x^*) - f(x^\circ) \leq d\frac{\epsilon r_{\text{opt}}}{4d\cdot\log m}H(\widetilde{y}) + \frac{d\frac{\epsilon r_{\text{opt}}}{2d\cdot\|x^*\|_2^2}}{2}\|x^*\|_2^2 \leq \frac{\epsilon r_{\text{opt}}}{2}$ using Claim 3.4. □

**Claim 3.6.** *For every $x \in \mathbb{R}^d$, we have $f(x^\circ) - f(x) \leq \|x - x^\circ\|_2$*

*Proof.* Slightly abusing notation, we denote by $\nabla f(x)$ any subgradient of $f(x)$ at point $x$. Or, in symbols, we write $\nabla f(x) = \widetilde{y}^T A$ for *any arbitrary* $\widetilde{y} \in \arg\min_{y\in\Delta_m}\phi(x^\circ, y)$. Since $\widetilde{y}^T A$ can be seen as a weighted combination of $A_1, A_2, \ldots, A_m$ because $\widetilde{y} \in \Delta_m$, we claim that $\|\nabla f(x)\|_2 \leq 1$ owing to the normalization that all $A_i$'s satisfy $\|A_i\|_2 \leq 1$. This is known as "$f(x)$ is 1-Lipschitz continuous" in the optimization language.

Next, using calculus we compute

$$f(x^\circ) - f(x) = \int_{\tau=0}^1 \langle \nabla f(x + \tau(x^\circ - x)), x^\circ - x\rangle d\tau \leq \int_{\tau=0}^1 \|x^\circ - x\|_2 d\tau = \|x^\circ - x\|_2\ . \quad\square$$

---

[3]The computation of a "warm start" $\beta$ is also needed by our immediate prior work [XSX06]; in Appendix C we include an optimization-based algorithm for computing $\beta$ faster and simpler than prior work.

**Algorithm 2** `MaxIBSPSolver`$(A, b, \varepsilon, \alpha, \beta)$

**Input:** Bounded polyhedron $P = \{x \colon Ax + b \geq 0\}$; error constant $\varepsilon$, aspect ratio upper bound $\alpha$, and $\beta$ satisfying $\beta/c \leq r_{\mathsf{opt}} \leq \beta$ for some constant $c \geq 1$.
**Output:** $x^{(T)}$ is an $1 - O(\varepsilon)$ approximation to MaxIB.
1: $\lambda \leftarrow \frac{\epsilon(\beta/c)}{4d \cdot \log m}$, $\gamma \leftarrow \frac{\epsilon}{2d \cdot 4\alpha^2 \beta}$, and $T \leftarrow \widetilde{\Omega}(d + \alpha\sqrt{d}/\varepsilon)$.
2: $x^{(T)}, y^{(T)} \leftarrow$ `L1L2SPSolver`$(A, b, \lambda, \gamma, T)$
3: **return** $x^{(T)}$.

**Our Algorithm.** We are now ready to apply `L1L2SPSolver` to our regularized saddle-point problem (2.3). Owing to Theorem 2.2, `L1L2SPSolver` produces a pair $x^{(T)}, y^{(T)}$ satisfying

$$\mathbb{E}[\|x^{(T)} - x^\circ\|_2] \leq \theta^{T/2} \cdot \left( (\frac{1}{\tau} + 2\lambda)\log m + (\frac{1}{2\sigma} + d\gamma)\|x^\circ\|_2^2 \right) \cdot \left( \frac{1}{4\sigma} + d\gamma \right)^{-1/2} ,$$

where

$$\theta = 1 - \frac{1}{d + q/\sqrt{\lambda d \gamma}} = 1 - \frac{1}{d + q/\sqrt{d \frac{\epsilon(\beta/c)}{4d \cdot \log m} \frac{\epsilon}{2d \cdot 4\alpha^2 \beta}}} = 1 - \frac{1}{d + \frac{\sqrt{32c \log m} \cdot q \alpha \sqrt{d}}{\varepsilon}} .$$

We choose

$$T \geq 2 \log_\theta \left( \frac{\epsilon r_{\mathsf{opt}}}{2} \cdot \left( \frac{1}{4\sigma} + d\gamma \right)^{1/2} \cdot \left( (\frac{1}{\tau} + 2\lambda)\log m + (\frac{1}{2\sigma} + d\gamma)\|x^\circ\|_2^2 \right)^{-1} \right) .$$

Since $q = O(\sqrt{\log m})$ and the formula inside the above $\log_\theta$ is within $\mathsf{poly}(m, d, \alpha, 1/\varepsilon)$, we use the $\widetilde{\Omega}$ to hide these logarithmic factors. In other words, after

$$T \geq \widetilde{\Omega}(1)/(1-\theta) = \widetilde{\Omega}(d + \sqrt{d}\alpha/\varepsilon)$$

iterations, we have $\mathbb{E}[\|x^{(T)} - x^\circ\|_2] \leq \frac{\epsilon r_{\mathsf{opt}}}{2}$. Combining this with Claim 3.5 and Claim 3.6, we claim that this output $x^{(T)}$ satisfies $r_{\mathsf{opt}} - \mathbb{E}[f(x^{(T)})] = \mathbb{E}[f(x^*) - f(x^{(T)})] \leq \varepsilon r_{\mathsf{opt}}$. Applying a simple Markov inequality, we have that with probability at least $2/3$, it satisfies $f(x^{(T)}) \geq (1 - 3\varepsilon)r_{\mathsf{opt}}$ as desired.

Finally, it is a simple exercise to show that each iteration of Algorithm 1 can be implemented to run in $O(m)$ time. Therefore, the total running time is $\widetilde{O}(md + m\sqrt{d}\alpha/\varepsilon)$. This finishes the proof of the following theorem:

> **Theorem 3.7.** *Suppose some value $\beta > 0$ is known and satisfies $\beta/c \leq r_{\mathsf{opt}} \leq \beta$ for some constant $c$. Then, `MaxIBSPSolver`$(A, b, \varepsilon, \alpha, \beta)$ produces a $(1 - 3\varepsilon)$ approximate solution to MaxIB with probability at least $2/3$. Furthermore, the total running time is*
> 
> $$\widetilde{O}\left(md + \frac{m\sqrt{d}\alpha}{\varepsilon}\right) .$$

Note that one can preprocess in time $\widetilde{O}(md + m\sqrt{d}\alpha)$ to obtain $\beta$, the constant approximation of $r_{\mathsf{opt}}$. Details of this preprocessing appear in Appendix C.

## 4 MinEB: From Geometry to Saddle-Point Optimization

In this section, we study the problem of finding the minimum enclosing ball (MinEB) of $n$ points in $d$ dimension space, using `L1L2SPSolver` proposed in Section 2. Below we first write MinEB as an optimization problem, and prove some basic facts in Section 4.1, then apply `L1L2SPSolver` in Section 4.2.

## 4.1 Optimization View of MinEB

Recall that the $n$ points are given as $\{a_1, a_2, ..., a_n\} \subseteq \mathbb{R}^d$, and we define the $d \times n$ matrix $A = [a_1, a_2, ..., a_n]$, where each $a_i$ is a column vector. Without loss of generality, we assume that $a_1 = 0$; if not, one can shift all the points and move the origin to $a_1$.[4] Also, without loss of generality, we can scale the matrix to satisfy $\max_{i \in [n]} \|a_i\|_2 = 1$ because we are interested in multiplicative approximations.

Consider the saddle-point problem

$$\mathsf{OPT} \stackrel{\text{def}}{=} \min_{y \in \mathbb{R}^d} \max_{x \in \Delta_n} \frac{1}{2} \sum_i x_i \|y - a_i\|_2^2 = \max_{x \in \Delta_n} \min_{y \in \mathbb{R}^d} \frac{1}{2} \sum_i x_i \|y - a_i\|_2^2 \ . \tag{4.1}$$

Strong duality holds for instance due to Sion's minimax theorem [Sio58].

It is clear by the definition of the saddle-point problem that $\mathsf{OPT}$ is equal to $\frac{1}{2} R_{\mathsf{opt}}^2$, where recall that $R_{\mathsf{opt}}$ is the radius of the minimum enclosing ball of the points. The following fact gives a lower bound on $\mathsf{OPT}$:

**Fact 4.1.** $1 \leq 8\mathsf{OPT}$.

*Proof.* Recall that $\mathsf{OPT} = \frac{1}{2} R_{\mathsf{opt}}^2$, where $R_{\mathsf{opt}}$ is the radius of the minimum enclosing ball. On the other hand, there exist a pair of points that are of distance 1 away from each other, since $\|a_i\|_2$ is the distance between point 1 and point $i$ (recall that $a_1 = 0$) and $\max_i \|a_i\|_2 = 1$. This further implies that, any enclosing ball of the given $n$ points must have radius at least $1/2$. In sum, we must have $1 \leq 2R_{\mathsf{opt}}$, which implies $1 \leq 8\mathsf{OPT}$. □

## 4.2 Apply `L1L2SPSolver` to MinEB

We rewrite (4.1) as

$$\mathsf{OPT} = \max_{x \in \Delta_n} \min_{y \in \mathbb{R}^d} \frac{1}{2} \sum_i x_i \|y - a_i\|_2^2 = -\max_{y \in \mathbb{R}^d} \min_{x \in \Delta_n} \left\{ \phi(x, y) \stackrel{\text{def}}{=} \phi(x, y) = x^T A^T y + x^T b - \frac{1}{2} \|y\|_2^2 \right\},$$

where $b_i \stackrel{\text{def}}{=} -\frac{1}{2} \|a_i\|_2^2$. In this section, we define $g(y) \stackrel{\text{def}}{=} \min_{x \in \Delta_n} \phi(x, y)$, and let $x^*, y^*$ be the optimal saddle point of this above saddle point problem. Note that $y^*$ is also necessarily a maximizer of $g(y)$ and $g(y^*) = -\mathsf{OPT}$. Now MinEB can be characterized as the following approximate maximization problem:

**Claim 4.2.** *Any $y$ satisfying $g(y) \geq g(y^*) - 2\varepsilon\mathsf{OPT}$ is a $(1 + \varepsilon)$-approximation to MinEB.*

*Proof.* The definition of $g$ tells us that $\sqrt{-2g(y)}$ is the minimum radius of the ball centered at $y$ enclosing all the given points. Therefore, it suffices to show that $\sqrt{-2g(y)} \leq (1+\varepsilon)R_{\mathsf{opt}}$. However, since we have $-2g(y) \leq -2g(y^*) + 4\varepsilon\mathsf{OPT} = 2(1+2\varepsilon)\mathsf{OPT} = (1+2\varepsilon)R_{\mathsf{opt}}^2$, taking the square root on both sides and using $\sqrt{1+2\varepsilon} \leq 1 + \varepsilon$ finish the proof. □

---
[4]This preprocessing of setting $a_1 = 0$ is used in proving Fact 4.1. We note also that one can use this preprocessing to improve the running time of the $\ell_1$ algorithm in [SVZ11] from $\widetilde{O}(ndL/\sqrt{\varepsilon})$ into $\widetilde{O}(nd/\sqrt{\varepsilon})$.

Unfortunately, $\phi(x,y)$ does not fall into the category of (2.1) because it is not strongly convex with respect to $x$. For this reason, define the regularized saddle-point problem

$$-\min_{x \in \Delta_n} \max_{y \in \mathbb{R}^d} \left\{ \phi_r(x,y) \stackrel{\text{def}}{=} x^T A^T y + x^T b + d\lambda H(x) - \frac{d\gamma}{2}\|y\|_2^2 \right\} , \qquad (4.2)$$

where the parameters

$$\lambda = \frac{\varepsilon}{8d \cdot \log n} \leq \frac{\varepsilon \mathsf{OPT}}{d \cdot \log n} \quad \text{and} \quad \gamma = \frac{1}{d} .$$

Above, the only inequality is owing to Fact 4.1. We denote by $(x^\circ, y^\circ)$ the optimal saddle-point of objective (4.2). We now make a few claims before we apply our saddle-point algorithm in Section 2.

**Claim 4.3.** $g(y^*) - g(y^\circ) \leq \varepsilon \mathsf{OPT}$.

*Proof.* Denoting $\widetilde{x} \stackrel{\text{def}}{=} \arg\min_{x \in \Delta_n} \phi(x, y^\circ)$, we have (by the definition of saddle points),

$$g(y^\circ) = \widetilde{x}^T A^T y^\circ + \widetilde{x}^T b - \frac{1}{2}\|y^\circ\|_2^2 = \phi_r(\widetilde{x}, y^\circ) - d\lambda H(\widetilde{x}) \geq \phi_r(x^\circ, y^\circ) - d\lambda H(\widetilde{x})$$
$$\geq \phi_r(x^\circ, y^*) - d\lambda H(\widetilde{x}) = \phi(x^\circ, y^*) - d\lambda H(\widetilde{x}) + d\lambda H(x^\circ)$$
$$\geq \phi(x^*, y^*) - d\lambda H(\widetilde{x}) + d\lambda H(x^\circ) = g(y^*) - d\lambda H(\widetilde{x}) + d\lambda H(x^\circ) \geq g(y^*) - d\lambda H(\widetilde{x})$$

Since $H(\widetilde{x}) \in [0, \log n]$, we have $g(y^*) - g(y^\circ) \leq d\lambda H(\widetilde{x}) \leq d\frac{\varepsilon \mathsf{OPT}}{d \cdot \log n} \log n \leq \varepsilon \mathsf{OPT}$. □

**Claim 4.4.** *For every $y \in \mathbb{R}^d$, we have $g(y^\circ) - g(y) \leq (1 + \|y^\circ\|_2)\|y^\circ - y\|_2$ .*

*Proof.* Slightly abusing notation, we denote by $\nabla g(y)$ *any* subgradient of $g(y)$ at point $y$. Or, in symbols, we write $\nabla g(y) = \widetilde{x}_y^T A^T - y$ for *any arbitrary* $\widetilde{x}_y \in \arg\min_{x \in \Delta_n} \phi(x, y^\circ)$. Since $\widetilde{x}_y^T A^T$ can be seen as a weighted combination of $a_1, a_2, \ldots, a_n$ because $\widetilde{x}_y \in \Delta_n$, we claim that $\|\widetilde{x}_y^T A^T\|_2 \leq 1$ owing to the normalization that all $a_i$'s satisfy $\|a_i\|_2 \leq 1$.

Next, using calculus we compute

$$g(y^\circ) - g(y) = \int_{\tau=0}^{1} \langle \nabla g(y + \tau(y^\circ - y)), y^\circ - y \rangle d\tau$$
$$= \int_{\tau=0}^{1} \langle \widetilde{x}_{y+\tau(y^\circ-y)}^T A^T - (y + \tau(y^\circ - y)), y^\circ - y \rangle d\tau$$
$$\leq \|\widetilde{x}_{y+\tau(y^\circ-y)}^T A^T\|_2 \|y^\circ - y\|_2 + \int_{\tau=0}^{1} \langle -y^\circ, y^\circ - y \rangle d\tau - \frac{\|y^\circ - y\|_2^2}{2}$$
$$\leq \|y^\circ - y\|_2 + \|y^\circ\|_2 \|y^\circ - y\|_2 = (1 + \|y^\circ\|_2)\|y^\circ - y\|_2 .$$ □

**Our Algorithm.** Now we are ready to apply `L1L2SPSolver` to our regularized saddle-point problem (4.2). Notice that we need to pass $A^T$ as the parameter $A$ to `L1L2SPSolver`, and we should treat $x$ as $y$ and $y$ as $x$ in the algorithm due to the difference between (4.2) and (2.1). We write this as `MinEBSPSolver` in Algorithm 3. Based on Theorem 2.2, `L1L2SPSolver` produces a pair $x^{(T)}, y^{(T)}$ satisfying

$$\mathbb{E}\left[\|y^{(T)} - y^\circ\|_2\right] \leq \theta^{T/2} \cdot \left( (\frac{1}{\tau} + 2\lambda) \log n + (\frac{1}{2\sigma} + d\gamma)\|y^\circ\|_2^2 \right) \cdot \left( \frac{1}{4\sigma} + d\gamma \right)^{-1/2} .$$

**Algorithm 3** `MinEBSPSolver((a_1,...,a_n), ε)`

**Input:** $n$ points $a_1, \ldots, a_n \in \mathbb{R}^d$ and error constant $\varepsilon > 0$
**Output:** $y^{(T)}$ is an $1 + O(\varepsilon)$ approximation to MinEB.
1: $A \leftarrow [a_1^T, \ldots, a_n^T] \in \mathbb{R}^{d \times n}$ and $b_i \leftarrow -\frac{1}{2}\|a_i\|_2^2$ for every $i \in [n]$.
2: $\lambda \leftarrow \frac{\varepsilon}{8d \cdot \log n}$, $\gamma \leftarrow \frac{1}{d}$, and $T \leftarrow \widetilde{\Omega}\big(d + \sqrt{\frac{d}{\varepsilon}}\big)$.
3: $y^{(T)}, x^{(T)} \leftarrow$ `L1L2SPSolver`$(A^T, b, \lambda, \gamma, T)$
4: **return** $y^{(T)}$.

The above inequality, together with Claim 4.4, tells us that in order to let $\mathbb{E}[g(y^\circ) - g(y^{(T)})] \leq \varepsilon \mathsf{OPT}$ it suffices to choose $T$ such that

$$\theta^{T/2} \cdot (1 + \|y^\circ\|_2) \left( (\frac{1}{\tau} + 2\lambda) \log n + (\frac{1}{2\sigma} + d\gamma) \|y^\circ\|_2^2 \right) \cdot \left( \frac{1}{4\sigma} + d\gamma \right)^{-1/2} \leq \varepsilon \mathsf{OPT} \ ,$$

where $\theta = 1 - \frac{1}{d + q/\sqrt{\lambda d\gamma}} = 1 - \frac{1}{d + q/\sqrt{\varepsilon/8d \log n}}$. In sum, we need to have

$$T \geq 2\log_\theta \left( \frac{\varepsilon \mathsf{OPT}}{1 + \|y^\circ\|_2} \cdot \left( \frac{1}{4\sigma} + d\gamma \right)^{1/2} \cdot \left( (\frac{1}{2\tau} + \lambda) \log n + (\frac{1}{2\sigma} + d\gamma) \|x^\circ\|_2^2 \right)^{-1} \right) \ ,$$

Since $q = O(\sqrt{\log n})$ and the formula inside the $\log_\theta$ is within $\mathsf{poly}(n, d, 1/\varepsilon)$, we can use the $\widetilde{\Omega}$ to hide these logarithmic factors. Thus, after $T \geq \widetilde{\Omega}(1)/(1-\theta) = \widetilde{\Omega}(d + \sqrt{d}/\sqrt{\varepsilon})$ iterations, we have $\mathbb{E}[g(y^\circ) - g(y^{(T)})] \leq \varepsilon \mathsf{OPT}$. Combining this with Claim 4.3, we know that this output $y^{(T)}$ satisfies $\mathbb{E}[g(y^*) - g(y^{(T)})] \leq 2\varepsilon \mathsf{OPT}$. Applying a simple Markov inequality, we have that with probability at least $2/3$, we have $g(y^*) - g(y^{(T)}) \leq 6\varepsilon \mathsf{OPT}$. In other words, with probability at least $2/3$, $y^{(T)}$ is a $(1 + 3\varepsilon)$-approximate solution to $R_{\mathsf{opt}}$ owing to Claim 4.2.

Finally, it is a simple exercise to show that each iteration of Algorithm 1 can be implemented to run in $O(n)$ time. Therefore, the total running time of `MinEBSPSolver` is $\widetilde{O}(nd + n\sqrt{d}/\sqrt{\varepsilon})$. This finishes the proof of the following theorem:

**Theorem 4.5.** `MinEBSPSolver`$((a_1, \ldots, a_n), \varepsilon)$ *produces a $(1 + 3\varepsilon)$ approximate solution to MinEB with probability at least $2/3$. Furthermore, the total running time is*

$$\widetilde{O}\Big(nd + \frac{n\sqrt{d}}{\sqrt{\varepsilon}}\Big) \ .$$

**Acknowledgements** We thank Lorenzo Orecchia for insightful comments and discussions. He took an active part in stages of this research, and yet declined to be a co-author. We thank Sepideh Mahabadi and Jinhui Xu for helpful conversations. We also thank the anonymous reviewers' valuable comments on the earlier versions of this paper.

# Appendix

## A  Proof of Theorem 2.2

In order to prove Theorem 2.2, we need the following lemmas.

**Lemma A.1.** *Let $x_2 = \arg\min_{z \in \Delta_m}\{\frac{V_{x_1}(z)}{\tau} + \langle \xi, z \rangle + \lambda H(z)\}$, then for every $u \in \Delta_m$, we have*

$$\frac{1}{\tau}V_{x_1}(u) - \left(\frac{1}{\tau} + \lambda\right)V_{x_2}(u) - \frac{1}{2\tau}\|x_2 - x_1\|_1^2 \geq \langle \xi, x_2 - u \rangle + \lambda H(x_2) - \lambda H(u)$$

*Proof.* By definition of $x_2$, we know for all $u \in \Delta_m$,

$$\langle \frac{\nabla V_{x_1}(x_2)}{\tau} + \xi + \lambda \nabla H(x_2), u - x_2 \rangle \geq 0 \Rightarrow \langle \frac{\nabla V_{x_1}(x_2)}{\tau} + \lambda \nabla H(x_2), u - x_2 \rangle \geq \langle \xi, x_2 - u \rangle \quad (A.1)$$

In addition, the following equation holds and is known as the three-point equality of Bregman divergence:

$$\left\langle \frac{\nabla V_{x_1}(x_2)}{\tau}, u - x_2 \right\rangle = \frac{1}{\tau}\langle \nabla H(x_2) - \nabla H(x_1), u - x_2 \rangle = \frac{1}{\tau}(V_{x_1}(u) - V_{x_2}(u) - V_{x_1}(x_2)) \ .$$

Also, recall that the definition of $V_{x_2}(u)$ tells us that

$$\lambda \langle \nabla H(x_2), u - x_2 \rangle = -\lambda(V_{x_2}(u) - H(u) + H(x_2)) \ .$$

Substituting the above two equalities into (A.1), we have

$$-\lambda(V_{x_2}(u) - H(u) + H(x_2)) + \frac{1}{\tau}(V_{x_1}(u) - V_{x_2}(u) - V_{x_1}(x_2)) \geq \langle \xi, x_2 - u \rangle$$
$$\Rightarrow \frac{1}{\tau}V_{x_1}(u) - \left(\frac{1}{\tau} + \lambda\right)V_{x_2}(u) - \frac{1}{2\tau}\|x_2 - x_1\|_1^2 \geq \langle \xi, x_2 - u \rangle + \lambda H(x_2) - \lambda H(u). \qquad \square$$

**Lemma A.2.** *Let $x = \arg\min_{z \in \Delta_m}\{z^T b + \lambda H(z)\}$, then for all $u \in \Delta_m$,*

$$u^T b + \lambda H(u) - x^T b - \lambda H(x) \geq \lambda V_x(u)$$

*Proof.* Define $g(z) \stackrel{\text{def}}{=} z^T b + \lambda H(z)$. By the minimality of $x$, we know that for all $u \in \Delta_m$ we must have $\langle \nabla g(x), u - x \rangle \geq 0$. Therefore, for all $u \in \Delta_m$,

$$g(u) - g(x) = u^T b + \lambda H(u) - x^T b - \lambda H(x)$$
$$\geq u^T b + \lambda H(u) - x^T b - \lambda H(x) - \langle \nabla g(x), u - x \rangle$$
$$= u^T b + \lambda H(u) - x^T b - \lambda H(x) - \langle \lambda \nabla H(x) + b, u - x \rangle$$
$$= \lambda H(u) - \lambda H(x) - \lambda \langle \nabla H(x), u - x \rangle = \lambda V_x(u)$$

$\square$

**Lemma A.3.** *The updating rules of $x^{(t+1)}$ and $y^{(t+1)}$ in* `L1L2SPSolver` *are equivalent to*

$$x_i^{(t+1)} = \begin{cases} \arg\min_x \left\{ -\langle y^{(t)} + \theta(y^{(t)} - y^{(t-1)}), A_i \rangle x_i + \frac{d\gamma}{2}x_i^2 + \frac{(x_i - x_i^{(t)})^2}{2\sigma} \right\} & \text{if } i = i^*, \\ x_i^{(t)} & \text{if } i \neq i^*. \end{cases}$$

$$y^{(t+1)} = \arg\min_{y \in \Delta_m} \left\{ \frac{1}{d}y^T A\left(x^{(t)} + d(x^{(t+1)} - x^{(t)})\right) + \frac{1}{d}y^T b + \lambda H(y) + \frac{V_{y^{(t)}}(y)}{\tau} \right\}$$

*Proof.* Easily verifiable by taking the gradient. The update rule on the $y$ side is also known as multiplicative weight update, see for instance [AO14]. □

We are now read to prove Theorem 2.2.

**Step I: Inequality for the $x$ side.** For every $i \in \{1, 2, \ldots, d\}$, define $\widetilde{x}_i$ to be the value of $x_i^{(t+1)}$ if $i = i^*$; or in symbols,

$$\widetilde{x}_i \stackrel{\text{def}}{=} \arg\min_{z \in \mathbb{R}} \left\{ -\langle y^{(t)} + \theta(y^{(t)} - y^{(t-1)}), A_i \rangle z + \frac{d\gamma}{2} z^2 + \frac{\|z - x_i^{(t)}\|_2^2}{2\sigma} \right\}$$

Since $\widetilde{x}_i$ is the minimizer, and since the function inside arg min is $d\gamma + \frac{1}{\sigma}$ strongly convex, we have[5]

$$- \langle y^{(t)} + \theta(y^{(t)} - y^{(t-1)}), A_i \rangle x_i^\circ + \frac{d\gamma}{2} x_i^{\circ 2} + \frac{(x_i^\circ - x_i^{(t)})^2}{2\sigma}$$
$$\geq - \langle y^{(t)} + \theta(y^{(t)} - y^{(t-1)}), A_i \rangle \widetilde{x}_i + \frac{d\gamma}{2} \widetilde{x}_i^2 + \frac{(\widetilde{x}_i - x_i^{(t)})_2^2}{2\sigma} + \left( \frac{1}{2\sigma} + \frac{d\gamma}{2} \right) (x_i^\circ - \widetilde{x}_i)_2^2 \ . \qquad \text{(A.2)}$$

On the other hand, by the definition of $x^\circ$, we have $x^\circ$ maximizes $y^{\circ T} Ax - \frac{\gamma d}{2} \|x\|_2^2$, and therefore $x_i^\circ$ maximizes $\langle y^\circ, A_i \rangle x_i - \frac{d\gamma x_i^2}{2}$, which is $d\gamma$ strongly concave with respect to $x_i$. This implies

$$\langle y^\circ, A_i \rangle x_i^\circ - \frac{d\gamma x_i^{\circ 2}}{2} \geq \langle y^\circ, A_i \rangle \widetilde{x}_i - \frac{d\gamma \widetilde{x}_i^2}{2} + \frac{d\gamma}{2} (\widetilde{x}_i - x_i^\circ)^2 \qquad \text{(A.3)}$$

Summing up (A.2) and (A.3), we get:

$$- \langle y^{(t)} + \theta(y^{(t)} - y^{(t-1)}), A_i \rangle x_i^\circ + \frac{(x_i^\circ - x_i^{(t)})^2}{2\sigma} + \langle y^\circ, A_i \rangle x_i^\circ$$
$$\geq - \langle y^{(t)} + \theta(y^{(t)} - y^{(t-1)}), A_i \rangle \widetilde{x}_i + \frac{(\widetilde{x}_i - x_i^{(t)})_2^2}{2\sigma} + \left( \frac{1}{2\sigma} + d\gamma \right) (x_i^\circ - \widetilde{x}_i)_2^2 + \langle y^\circ, A_i \rangle \widetilde{x}_i$$

After simplification, this is

$$\frac{(x_i^\circ - x_i^{(t)})^2}{2\sigma} \geq \langle y^{(t)} + \theta(y^{(t)} - y^{(t-1)}) - y^\circ, A_i \rangle (x_i^\circ - \widetilde{x}_i) + \frac{(\widetilde{x}_i - x_i^{(t)})_2^2}{2\sigma} + \left( \frac{1}{2\sigma} + d\gamma \right) (x_i^\circ - \widetilde{x}_i)_2^2 \qquad \text{(A.4)}$$

Let $\mathcal{F}_t$ be the sigma field generated by all random variables defined before round $t$, and taking expectation conditioned on $\mathcal{F}_t$, we have

$$\mathbb{E}[x_i^{(t+1)} | \mathcal{F}_t] = \frac{1}{d} \widetilde{x}_i + \frac{d-1}{d} x_i^{(t)}$$
$$\mathbb{E}\left[ (x_i^\circ - x_i^{(t+1)})^2 | \mathcal{F}_t \right] = \frac{1}{d}(x_i^\circ - \widetilde{x}_i)^2 + \frac{d-1}{d}(x_i^\circ - x_i^{(t)})^2$$
$$\mathbb{E}\left[ (x_i^{(t+1)} - x_i^{(t)})^2 | \mathcal{F}_t \right] = \frac{1}{d}(\widetilde{x}_i - x_i^{(t)})^2$$

So we know:

$$\widetilde{x}_i = -(d-1)x_i^{(t)} + d\mathbb{E}\left[x_i^{(t+1)} | \mathcal{F}_t\right] = x_i^{(t)} + d\mathbb{E}\left[x_i^{(t+1)} - x_i^{(t)} | \mathcal{F}_t\right]$$

---
[5]Indeed, if $g(z)$ is $\zeta$ strongly convex and $\widetilde{z} = \arg\min_z \{g(z)\}$, then $g(z) - g(\widetilde{z}) \geq \frac{\zeta}{2}(z - \widetilde{z})^2$ for all $z$.

$$(x_i^\circ - \widetilde{x}_i)^2 = -(d-1)(x_i^\circ - x_i^{(t)})^2 + d\mathbb{E}\left[(x_i^\circ - x_i^{(t+1)})^2 | \mathcal{F}_t\right]$$

Substituting them into (A.4), we get:

$$\frac{(x_i^\circ - x_i^{(t)})^2}{2\sigma} \geq \langle y^{(t)} + \theta(y^{(t)} - y^{(t-1)}) - y^\circ, A_i\rangle \left(x_i^\circ - x_i^{(t)} - d\mathbb{E}\left[x_i^{(t+1)} - x_i^{(t)} | \mathcal{F}_t\right]\right) +$$

$$\frac{d\mathbb{E}\left[(x_i^{(t+1)} - x_i^{(t)})^2 | \mathcal{F}_t\right]}{2\sigma} + \left(\frac{1}{2\sigma} + d\gamma\right)\left(-(d-1)(x_i^\circ - x_i^{(t)})^2 + d\mathbb{E}\left[(x_i^\circ - x_i^{(t+1)})^2 | \mathcal{F}_t\right]\right)$$

Or equivalently,

$$\left(\frac{d}{2\sigma} + (d-1)d\gamma\right)(x_i^\circ - x_i^{(t)})^2 \geq \langle y^{(t)} + \theta(y^{(t)} - y^{(t-1)}) - y^\circ, A_i\rangle \left(x_i^\circ - x_i^{(t)} - d\mathbb{E}\left[x_i^{(t+1)} - x_i^{(t)} | \mathcal{F}_t\right]\right) +$$

$$\frac{d\mathbb{E}\left[(x_i^{(t+1)} - x_i^{(t)})^2 | \mathcal{F}_t\right]}{2\sigma} + \left(\frac{d}{2\sigma} + d^2\gamma\right)\mathbb{E}\left[(x_i^\circ - x_i^{(t+1)})^2 | \mathcal{F}_t\right]$$

Summing over all $i$ and divide by $d$, we get

$$\left(\frac{1}{2\sigma} + (d-1)\gamma\right)\|x^\circ - x^{(t)}\|_2^2 \geq \left\langle y^{(t)} + \theta(y^{(t)} - y^{(t-1)}) - y^\circ, -u^{(t)} + u^\circ - \sum_i A_i^T(\mathbb{E}[x_i^{(t+1)} - x_i^{(t)} | \mathcal{F}_t])\right\rangle +$$

$$\frac{\mathbb{E}\left[\|x^{(t+1)} - x^{(t)}\|_2^2 | \mathcal{F}_t\right]}{2\sigma} + \left(\frac{1}{2\sigma} + d\gamma\right)\mathbb{E}\left[\|x^\circ - x^{(t+1)}\|_2^2 | \mathcal{F}_t\right] \quad (A.5)$$

Here, $u^\circ = \frac{1}{d}\sum_i A_i^T x_i^\circ$, and $u^{(t)} = \frac{1}{d}\sum_i A_i^T x_i^{(t)}$.

**Step II: Inequality for the $y$ side.** Based on Lemma A.1, choosing $u = y^\circ$, we know

$$\lambda H(y^\circ) + \frac{V_{y^{(t)}}(y^\circ)}{\tau} - \left(\frac{1}{\tau} + \lambda\right)V_{y^{(t+1)}}(y^\circ) - \frac{\|y^{(t)} - y^{(t+1)}\|_1^2}{2\tau}$$

$$\geq \frac{1}{d}(y^{(t+1)} - y^\circ)^T A\left(x^{(t)} + d(x^{(t+1)} - x^{(t)})\right) + \frac{1}{d}(y^{(t+1)} - y^\circ)^T b + \lambda H(y^{(t+1)}) \ .$$

On the other hand, since $y^\circ$ minimize $y^T A x^\circ + y^T b + d\lambda H(y)$, by Lemma A.2, we have

$$\frac{1}{d}y^{(t+1)T} A x^\circ + \frac{1}{d}y^{(t+1)T} b + \lambda H(y^{(t+1)}) = \frac{1}{d}y^{\circ T} A x^\circ + \frac{1}{d}y^{\circ T} b + \lambda H(y^\circ) + \lambda V_{y^{(t+1)}}(y^\circ) \ .$$

Summing them up, we have:

$$\frac{V_{y^{(t)}}(y^\circ)}{\tau} \geq \frac{1}{d}(y^{(t+1)} - y^\circ)^T A\left(x^{(t)} + d(x^{(t+1)} - x^{(t)}) - x^\circ\right) + \frac{\|y^{(t)} - y^{(t+1)}\|_1^2}{2\tau} + \left(\frac{1}{\tau} + 2\lambda\right)V_{y^{(t+1)}}(y^\circ)$$

$$\quad (A.6)$$

**Step III: Putting It All Together.** Take expectation and then take summation of (A.5) and (A.6), we have

$$\left(\frac{1}{2\sigma} + (d-1)\gamma\right)\|x^\circ - x^{(t)}\|_2^2 + \frac{V_{y^{(t)}}(y^\circ)}{\tau}$$

$$
\geq \mathbb{E}\left[\overbrace{\left(y^{(t+1)} - y^{(t)} - \theta(y^{(t)} - y^{(t-1)})\right)^T A \left(\frac{x^{(t)} - x^\circ}{d} + x^{(t+1)} - x^{(t)}\right)}^{(\star)} \bigg| \mathcal{F}_t\right]
$$
$$
+ \frac{\mathbb{E}\left[\|x^{(t+1)} - x^{(t)}\|_2^2 | \mathcal{F}_t\right]}{2\sigma} + \left(\frac{1}{2\sigma} + d\gamma\right) \mathbb{E}\left[\|x^\circ - x^{(t+1)}\|_2^2 | \mathcal{F}_t\right]
$$
$$
+ \mathbb{E}\left[\frac{\|y^{(t)} - y^{(t+1)}\|_1^2}{2\tau} \bigg| \mathcal{F}_t\right] + \mathbb{E}\left[\left(\frac{1}{\tau} + 2\lambda\right) V_{y^{(t+1)}}(y^\circ) \bigg| \mathcal{F}_t\right] . \tag{A.7}
$$

Now, observe that

$$
(\star) = \frac{1}{d}(y^{(t+1)} - y^{(t)})^T A(x^{(t+1)} - x^\circ) - \frac{\theta}{d}(y^{(t)} - y^{(t-1)})^T A(x^{(t)} - x^\circ)
$$
$$
+ \frac{d-1}{d}(y^{(t+1)} - y^{(t)})^T A(x^{(t+1)} - x^{(t)}) - \theta(y^{(t)} - y^{(t-1)})^T A(x^{(t+1)} - x^{(t)}) \tag{A.8}
$$

We now bound the right hand side of (A.8) as follows. First,

$$
(y^{(t+1)} - y^{(t)})^T A(x^{(t+1)} - x^{(t)})
$$
$$
\overset{\text{①}}{\geq} -\frac{\|(y^{(t)} - y^{(t+1)})\|_1^2}{4\tau} - \frac{\|A(x^{(t)} - x^{(t+1)})\|_\infty^2}{1/\tau}
$$
$$
\overset{\text{②}}{\geq} -\frac{\|(y^{(t)} - y^{(t+1)})\|_1^2}{4\tau} - \frac{q^2\|(x^{(t)} - x^{(t+1)})\|_\infty^2}{d/\tau}
$$
$$
\overset{\text{③}}{=} -\frac{\|(y^{(t)} - y^{(t+1)})\|_1^2}{4\tau} - \frac{\|(x^{(t)} - x^{(t+1)})\|_\infty^2}{4\sigma}
$$
$$
\overset{\text{④}}{=} -\frac{\|(y^{(t)} - y^{(t+1)})\|_1^2}{4\tau} - \frac{\|(x^{(t)} - x^{(t+1)})\|_2^2}{4\sigma} .
$$

Above, ① uses the inequality $\langle a, b \rangle \leq \|a\|_1 \|b\|_\infty \leq \frac{\|a\|_1^2}{2} + \frac{\|b\|_\infty^2}{2}$; ② uses the fact that $x^{(t)} - x^{(t+1)}$ is only a singleton vector (because $x$ changes at most one coordinate per iteration) as well as the assumption that the absolute values of all entries of $A$ are at most $q/\sqrt{d}$; ③ uses our choice of $\sigma\tau = \frac{d}{4q^2}$, and ④ again uses the fact that $x^{(t)} - x^{(t+1)}$ is a singleton.

Similarly, we also have

$$
-(y^{(t)} - y^{(t-1)})^T A(x^{(t+1)} - x^{(t)}) \geq -\frac{\|y^{(t)} - y^{(t-1)}\|_1^2}{4\tau} - \frac{\|A(x^{(t)} - x^{(t+1)})\|_\infty^2}{1/\tau}
$$
$$
\geq -\frac{\|y^{(t)} - y^{(t-1)}\|_1^2}{4\tau} - \frac{\|x^{(t)} - x^{(t+1)}\|_2^2}{4\sigma}
$$

Combining these with (A.8) and (A.7), we have:

$$
\left(\frac{1}{2\sigma} + (d-1)\gamma\right) \|x^\circ - x^{(t)}\|_2^2 + \frac{V_{y^{(t)}}(y^\circ)}{\tau}
$$
$$
\geq \mathbb{E}\bigg[\frac{1}{d}(y^{(t+1)} - y^{(t)})^T A(x^{(t+1)} - x^\circ) - \frac{\theta}{d}(y^{(t)} - y^{(t-1)})^T A(x^{(t)} - x^\circ)
$$
$$
+ \frac{d-1}{d}\left(-\frac{\|(y^{(t)} - y^{(t+1)})\|_1^2}{4\tau} - \frac{\|(x^{(t)} - x^{(t+1)})\|_2^2}{4\sigma}\right) - \theta\left(\frac{\|y^{(t)} - y^{(t-1)}\|_1^2}{4\tau} + \frac{\|(x^{(t)} - x^{(t+1)})\|_2^2}{4\sigma}\right) \bigg| \mathcal{F}_t\bigg]
$$

$$+ \frac{\mathbb{E}\left[\|x^{(t+1)} - x^{(t)}\|_2^2 | \mathcal{F}_t\right]}{2\sigma} + \left(\frac{1}{2\sigma} + d\gamma\right) \mathbb{E}\left[\|x^\circ - x^{(t+1)}\|_2^2 | \mathcal{F}_t\right]$$

$$+ \frac{\mathbb{E}\left[\|y^{(t)} - y^{(t+1)}\|_1^2 | \mathcal{F}_t\right]}{2\tau} + \left(\frac{1}{\tau} + 2\lambda\right) \mathbb{E}\left[V_{y^{(t+1)}}(y^\circ) | \mathcal{F}_t\right]$$

Note that we have $\frac{d-1}{d} \leq 1$ and we have chosen $\theta \leq 1$. Therefore, the above inequality implies that

$$\left(\frac{1}{2\sigma} + (d-1)\gamma\right) \|x^\circ - x^{(t)}\|_2^2 + \frac{V_{y^{(t)}}(y^\circ)}{\tau}$$

$$\geq \left(\frac{1}{\tau} + 2\lambda\right) \mathbb{E}\left[V_{y^{(t+1)}}(y^\circ) | \mathcal{F}_t\right] + \left(\frac{1}{2\sigma} + d\gamma\right) \mathbb{E}\left[\|x^\circ - x^{(t+1)}\|_2^2 | \mathcal{F}_t\right]$$

$$+ \frac{\mathbb{E}\left[\|y^{(t)} - y^{(t+1)}\|_1^2 | \mathcal{F}_t\right]}{4\tau} - \frac{\theta \|y^{(t)} - y^{(t-1)}\|_1^2}{4\tau}$$

$$+ \frac{1}{d}\mathbb{E}\left[(y^{(t+1)} - y^{(t)})^T A(x^{(t+1)} - x^\circ) | \mathcal{F}_t\right] - \frac{\theta}{d}(y^{(t)} - y^{(t-1)})^T A(x^{(t)} - x^\circ) \quad \text{(A.9)}$$

Finally, defining $\Delta^{(t)}$ to be

$$\Delta^{(t)} \overset{\text{def}}{=} \left(\frac{1}{\tau} + 2\lambda\right) \mathbb{E}\left[V_{y^{(t)}}(y^\circ) | \mathcal{F}_t\right] + \left(\frac{1}{2\sigma} + d\gamma\right) \mathbb{E}\left[\|x^\circ - x^{(t)}\|_2^2 | \mathcal{F}_t\right]$$

$$+ \frac{\|y^{(t)} - y^{(t-1)}\|_1^2}{4\tau} + \frac{1}{d}(y^{(t)} - y^{(t-1)})^T A(x^{(t)} - x^\circ) \ ,$$

we claim that (A.9) implies $\Delta^{(t+1)} \leq \theta \cdot \Delta^{(t)}$. This is in fact because our parameter choices of $\tau = \frac{1}{2q}\sqrt{\frac{d\gamma}{\lambda}}$, $\sigma = \frac{1}{2q}\sqrt{\frac{d\lambda}{\gamma}}$, and $\theta = 1 - \frac{1}{d+q/\sqrt{\lambda d\gamma}}$ together imply that

$$\frac{1/\tau}{1/\tau + 2\lambda} = 1 - \frac{1}{1 + 1/(2\tau\lambda)} = 1 - \frac{1}{1 + q/\sqrt{d\lambda\gamma}} \leq \theta \ , \text{and}$$

$$\frac{1/(2\sigma) + (d-1)\gamma}{1/(2\sigma) + d\gamma} = 1 - \frac{1}{d + 1/(2\sigma\gamma)} = 1 - \frac{1}{d + q/\sqrt{d\lambda\gamma}} = \theta \ .$$

In sum, we have $\Delta^{(t+1)} \leq \theta \cdot \Delta^{(t)}$ and also $\Delta^{(t)} \leq \theta^t \cdot \Delta^{(0)}$ where the boundary $\Delta^{(0)} = \left(\frac{1}{\tau} + 2\lambda\right) V_{y^{(0)}}(y^\circ) + \left(\frac{1}{2\sigma} + d\gamma\right) \|x^\circ - x^{(0)}\|_2^2$ because $y^{(-1)} = y^{(0)}$.

On the other hand, we also claim that

$$\Delta^{(t)} \geq \left(\frac{1}{\tau} + 2\lambda\right) \mathbb{E}\left[V_{y^{(t)}}(y^\circ) | \mathcal{F}_t\right] + \left(\frac{1}{4\sigma} + d\gamma\right) \mathbb{E}\left[\|x^\circ - x^{(t)}\|_2^2 | \mathcal{F}_t\right] \ . \quad \text{(A.10)}$$

It is easy to verify that (A.10) is a direct consequence of the definition of $\Delta^{(t)}$ as well as the following inequality.

$$\frac{1}{d}(y^{(t)} - y^{(t-1)})^T A(x^{(t)} - x^\circ) \overset{\text{\textcircled{1}}}{\geq} -\frac{\|(y^{(t)} - y^{(t-1)})\|_1^2}{4\tau} - \frac{\|A(x^{(t)} - x^\circ)\|_\infty^2}{d^2/\tau}$$

$$\overset{\text{\textcircled{2}}}{=} -\frac{\|(y^{(t)} - y^{(t-1)})\|_1^2}{4\tau} - \frac{\max_i(\langle A_i, x^{(t)} - x^\circ\rangle)^2}{d^2/\tau} \overset{\text{\textcircled{3}}}{\geq} -\frac{\|(y^{(t)} - y^{(t-1)})\|_1^2}{4\tau} - \frac{\|(x^{(t)} - x^\circ)\|_2^2}{d^2/\tau}$$

$$\overset{\text{\textcircled{4}}}{\geq} -\frac{\|(y^{(t)} - y^{(t-1)})\|_1^2}{4\tau} - \frac{\|(x^{(t)} - x^\circ)\|_2^2}{4\sigma} \ .$$

Above, ① uses the inequality $\langle a, b\rangle \leq \|a\|_1 \|b\|_\infty \leq \frac{\|a\|_1^2}{2} + \frac{\|b\|_\infty^2}{2}$; ② uses the definition of the $\ell_\infty$ norm; ③ uses the fact that $\|A_i\|_2 \leq 1$ according to our assumption on $A$; and ④ uses our choice of $\sigma\tau = \frac{d}{4q^2}$.

In sum, combining (A.10) and the just proved fact that $\Delta^{(T)} \leq \theta^T \cdot \Delta^{(0)}$ we have

$$\left(\frac{1}{\tau} + 2\lambda\right) \mathbb{E}\left[V_{y^{(T)}}(y^\circ)\right] + \left(\frac{1}{4\sigma} + d\gamma\right) \mathbb{E}[\|x^\circ - x^{(T)}\|_2^2]$$
$$\leq \theta^T \cdot \left(\left(\frac{1}{\tau} + 2\lambda\right) V_{y^{(0)}}(y^\circ) + \left(\frac{1}{2\sigma} + d\gamma\right) \|x^\circ - x^{(0)}\|_2^2\right)$$
$$\overset{①}{\leq} \theta^T \cdot \left(\left(\frac{1}{\tau} + 2\lambda\right) \log m + \left(\frac{1}{2\sigma} + d\gamma\right) \|x^\circ\|_2^2\right)$$

Above, ① holds because $V_{y^{(0)}}(y^\circ) \leq \log m$ due to our choice of $y^{(0)} = (1/m, \ldots, 1/m)$. □

## B  MaxIB: From Geometry to Convex Smooth Optimization

In this section we rewrite our saddle-point formulation of MaxIB in Lemma 3.2 as convex smooth optimization. Given some smoothing parameter $\mu > 0$ to be specified later, let us denote by

$$f(x) \overset{\text{def}}{=} \min_{y \in \Delta_m} y^T(Ax + b) \quad \text{and} \quad f_\mu(x) \overset{\text{def}}{=} \min_{y \in \Delta_m} \{y^T(Ax + b) + \mu H(y)\}, \tag{B.1}$$

where $H(y) = \log m + \sum_{i=1}^m y_i \log(y_i) \in [0, \log m]$ is the entropy function defined over the simplex $\Delta_m$. We remark here that, in some literature, $f_\mu(x)$ is known as the "soft min" over the simplex so is a natural smoothing variant of $f(x)$.

It is clear from Lemma 3.2 that we want to maximize $f(x)$ over $x \in \mathbb{R}^d$, but in fact, this maximization is approximately equivalent to the maximization on $f_\mu(x)$ (see Proposition B.1.a), and this new function $f_\mu(x)$ satisfies some smooth property (see Proposition B.1.d). We state the following simple properties about $f_\mu(x)$ and include their proofs only for completeness' sake.

**Proposition B.1.**
(a) For every $x \in \mathbb{R}^d$, we have $f(x) \leq f_\mu(x) \leq f(x) + \mu \log m$.
(b) For every $x \in \mathbb{R}^d$, letting $p(x) \overset{\text{def}}{=} \arg\min_{y \in \Delta_m} \{y^T(Ax + b) + \mu H(y)\}$, we have

$$p_j(x) = \frac{\exp^{-\frac{1}{\mu}(\langle A_j, x\rangle + b)}}{\sum_{j' \in [m]} \exp^{-\frac{1}{\mu}(\langle A_{j'}, x\rangle + b)}} \quad \text{and} \quad f_\mu(x) = \mu \log m - \mu \log \sum_{j \in [m]} \exp^{-\frac{1}{\mu}(\langle A_j, x\rangle + b)}.$$
(B.2)
(c) $f_\mu(x)$ is concave and continuously differentiable; it satisfies $\nabla f_\mu(x) = A^T p(x)$ for every $x \in \mathbb{R}^d$.
(d) For every $x_1, x_2 \in \mathbb{R}^d$, we have $\|\nabla f_\mu(x_1) - \nabla f_\mu(x_2)\|_2 \leq \frac{1}{\mu} \cdot \|x_1 - x_2\|_2$.

Notice that in the optimization language, Proposition B.1.d is known as $f_\mu(\cdot)$ being $\frac{1}{\mu}$-smooth with respect to the Euclidean norm [Nes04].

*Proof.*
(a) For every $y \in \Delta_m$, we have

$$y^T(Ax + b) \leq y^T(Ax + b) + \mu H(y) \leq y^T(Ax + b) + \mu \log m.$$

Next, taking minimization over $y \in \Delta_m$ immediately gives the desired inequality.

(b) This can be computed easily by taking the derivative.
(c) This can be easily verified from the explicit form of $f_\mu(x)$ in (B.2).
(d) Recall that if $a^* = \arg\min_{a \in \Delta_m} \phi(a)$ for some convex and differentiable function $\phi(\cdot)$, the minimality condition tells us that $\langle \nabla\phi(a^*), a^* - a \rangle \leq 0$ for all $a \in \Delta_m$. As a consequence, since $p(x_1) \stackrel{\text{def}}{=} \arg\min_{y \in \Delta_m}\{y^T(Ax_1+b) + \mu H(y)\}$ and $p(x_2) \stackrel{\text{def}}{=} \arg\min_{y \in \Delta_m}\{y^T(Ax_2+b) + \mu H(y)\}$, we have

$$\langle Ax_1 + b + \mu \nabla H(p(x_1)),\ p(x_1) - p(x_2) \rangle \leq 0\ ,\ \text{and}$$
$$\langle Ax_2 + b + \mu \nabla H(p(x_2)),\ p(x_2) - p(x_1) \rangle \leq 0\ .$$

Adding them together, we have:

$$\langle A(x_1 - x_2) + \mu \nabla H(p(x_1)) - \mu \nabla H(p(x_2)),\ p(x_1) - p(x_2) \rangle \leq 0$$
$$\iff \mu \langle \nabla H(p(x_1)) - \nabla H(p(x_2)),\ p(x_1) - p(x_2) \rangle \leq \langle A^T(p(x_1) - p(x_2)),\ x_1 - x_2 \rangle\ . \quad \text{(B.3)}$$

Now, since $H(\cdot)$ is 1-strongly convex over $\Delta_m$ with respect to the $\ell_1$-norm,[6] the left hand side of (B.3) is lower bounded as

$$\|p(x_1) - p(x_2)\|_1^2 \leq \langle \nabla H(p(x_1)) - \nabla H(p(x_2)),\ p(x_1) - p(x_2) \rangle\ . \quad \text{(B.4)}$$

Therefore, we deduce the following sequence of inequalities

$$\|A^T p(x_1) - A^T p(x_2)\|_2^2 \stackrel{①}{\leq} \|A\|_{2,\infty}^2 \|p(x_1) - p(x_2)\|_1^2 \stackrel{②}{\leq} \|p(x_1) - p(x_2)\|_1^2$$
$$\stackrel{③}{\leq} \frac{1}{\mu}\langle A^T(p(x_1) - p(x_2)), x_1 - x_2 \rangle$$
$$\stackrel{④}{\leq} \frac{1}{\mu}\|A^T p(x_1) - A^T p(x_2)\|_2 \|x_1 - x_2\|_2\ .$$

Above, ① uses the matrix $2-\infty$ norm $\|A\|_{2,\infty} \stackrel{\text{def}}{=} \max_x\{\max_{1 \leq j \leq m} |\langle A_j, x \rangle|\ :\ \|x\|_2 = 1\}$, ② uses the fact that $|\langle A_j, x \rangle| \leq \sqrt{\sum_i A_{ij}^2}\|x\|_2 = \|x\|_2$ which implies $\|A\|_{2,\infty} \leq 1$, ③ uses (B.3) and (B.4), and ④ uses the Cauchy-Schwarz inequality. Finally, dividing both sides by $\|A^T p(x_1) - A^T p(x_2)\|_2$ gives the desired inequality. □

**Our Algorithm.** Since our objective $f_\mu(\cdot)$ is concave and $\frac{1}{\mu}$-smooth with respect to the Euclidean norm, this is a good place to apply the *accelerated gradient method* of Nesterov (see [Nes83, Nes04, Nes05b]), which is the optimal first-order method for the class of smooth functions. We write our algorithm `MaxIBConvexSolver` in Algorithm 4.

Unfortunately, for technical reasons, existing results on accelerated gradient methods do not directly apply in our setting. Indeed, if the output of `MaxIBConvexSolver` is $\mathsf{w}^* \in \mathbb{R}^d$, the original convergence statement from accelerated gradient method states that $f_\mu(u) - f_\mu(\mathsf{w}^*) \leq f_\mu(u^*) - f_\mu(\mathsf{w}^*) \leq O(\|u^*\|_2^2/\mu T^2)$ for every $u \in \mathbb{R}^d$, where $u^*$ is the maximizer of $f_\mu(u)$. However, for our purpose of this paper, we need to turn the right-hand side into $O(\|u\|_2^2/\mu T^2)$. Since we cannot find this result stated anywhere in the literature, we prove it by making some routine changes to the classical proof. More precisely, we state the following theorem.

---

[6]This is a commonly known fact in optimization literature, see for instance [Sha11]. One of the equivalent definitions of a 1-strongly convex function $r(\cdot)$ is to demand that $\langle \nabla r(x) - \nabla r(y), x - y \rangle \geq \|x - y\|^2$.

**Algorithm 4** `MaxIBConvexSolver`$(f_\mu, T)$

**Input:** $f_\mu$ the smoothed objective defined in (B.1), and $T$ the number of iterations.
**Output:** $\mathsf{w}^*$ such that $f_\mu(u) - f_\mu(\mathsf{w}^*) \leq \frac{2\|u\|_2^2}{\mu T^2}$ for all $u \in \mathbb{R}^d$.

1: $L \leftarrow \frac{1}{\mu}$.
2: $\mathsf{v}_0 = \mathsf{w}_0 = \mathsf{z}_0 \leftarrow 0$ and $\mathsf{w}^* \leftarrow 0$.
3: **for** $k \leftarrow 0$ **to** $T-1$ **do**
4: $\quad \alpha_{k+1} \leftarrow \frac{k+2}{2L}$, and $\tau_k \leftarrow \frac{1}{\alpha_{k+1}L} = \frac{2}{k+2}$.
5: $\quad \mathsf{v}_{k+1} \leftarrow \tau_k \mathsf{z}_k + (1-\tau_k)\mathsf{w}_k$.
6: $\quad$ Compute $p(\mathsf{v}_{k+1})$ according to (B.2): $p_j(\mathsf{v}) = \frac{\exp^{-\frac{1}{\mu}(\langle A_j, \mathsf{v}\rangle + b)}}{\sum_{j'\in[m]}\exp^{-\frac{1}{\mu}(\langle A_{j'}, \mathsf{v}\rangle + b)}}$.
7: $\quad \mathsf{w}_{k+1} \leftarrow \mathsf{v}_{k+1} + \frac{1}{L}A^T p(\mathsf{v}_{k+1})$.
8: $\quad \mathsf{z}_{k+1} \leftarrow \mathsf{z}_k + \alpha_{k+1}A^T p(\mathsf{v}_{k+1})$.
$\qquad\qquad\qquad\diamond$ *we have* $\mathsf{w}_{k+1} = \arg\min_{\mathsf{w}}\left\{\frac{L}{2}\|\mathsf{w} - \mathsf{v}_{k+1}\|_2^2 - \langle\nabla f_\mu(\mathsf{v}_{k+1}), \mathsf{w} - \mathsf{v}_{k+1}\rangle\right\}$
$\qquad\qquad\qquad\diamond$ *we have* $\mathsf{z}_{k+1} = \arg\min_{\mathsf{z}}\left\{\frac{1}{2}\|\mathsf{z} - \mathsf{z}_k\|_2^2 - \langle\alpha_{k+1}\nabla f_\mu(\mathsf{v}_{k+1}), \mathsf{z} - \mathsf{z}_k\rangle\right\}$
9: $\quad$ **if** $f_\mu(\mathsf{w}_{k+1}) > f_\mu(\mathsf{w}^*)$ **then** $\mathsf{w}^* \leftarrow \mathsf{w}_{k+1}$ **end if**
10: **end for**
11: **return** $\mathsf{w}^*$.

**Theorem B.2.** *If $f_\mu(x)$ is concave and $\frac{1}{\mu}$-smooth with respect to $\|\cdot\|_2$, then* `MaxIBConvexSolver`$(f_\mu, T)$ *ensures*

$$\forall u \in \mathbb{R}^d, \qquad f_\mu(u) - f_\mu(\mathsf{w}^*) \leq \frac{2\|u\|_2^2}{\mu T^2}.$$

*Proof.* (There are multiple versions of the accelerated gradient method, and our `MaxIBConvexSolver` is written from the version provided in [AO14].)

Since $f_\mu(x)$ is concave and $-f_\mu(x)$ is convex, we apply [AO14, Lemma 4.3], the key lemma of the classical convergence proof of the accelerated gradient method, and obtain that for every $u \in Q$,

$$(\alpha_{k+1}^2 L)(-f_\mu(\mathsf{w}_{k+1})) - (\alpha_{k+1}^2 L - \alpha_{k+1})(-f_\mu(\mathsf{w}_k)) + \left(\frac{1}{2}\|\mathsf{z}_{k+1} - u\|_2^2 - \frac{1}{2}\|\mathsf{z}_k - u\|_2^2\right)$$
$$\leq \alpha_{k+1}(-f_\mu(u)).$$

Since our choice of $\alpha_k = \frac{k+1}{2L}$ ensures that $\alpha_k^2 L = \alpha_{k+1}^2 L - \alpha_{k+1} + \frac{1}{4L}$, we can telescope the above inequality with $k = 0, 1, \ldots, T-1$ and deduce that

$$\alpha_T^2 L(-f_\mu(\mathsf{w}_T)) + \sum_{k=1}^{T-1}\frac{1}{4L}(-f_\mu(\mathsf{w}_k)) + \left(\frac{1}{2}\|\mathsf{z}_T - u\|_2^2 - \frac{1}{2}\|\mathsf{z}_0 - u\|_2^2\right) \leq \sum_{k=1}^{T}\alpha_k(-f_\mu(u)).$$

Using $\sum_{k=1}^T \alpha_k = \frac{T(T+3)}{4L} = \alpha_T^2 L + \frac{T-1}{4L}$, $-f_\mu(\mathsf{w}^*) \leq -f_\mu(\mathsf{w}_k)$ (by Line 9 of `MaxIBConvexSolver`), and $\mathsf{z}_0 = 0$, we derive that

$$\frac{T(T+3)}{4L}(-f_\mu(\mathsf{w}^*)) \leq \alpha_T^2 L(-f_\mu(\mathsf{w}_T)) + \sum_{k=1}^{T-1}\frac{1}{4L}(-f_\mu(\mathsf{w}_k)) \leq \frac{T(T+3)}{4L}(-f_\mu(u)) + \frac{1}{2}\|u\|_2^2,$$

which after simplification implies $-f_\mu(\mathsf{w}_T) \leq (-f_\mu(u)) + \frac{2\|u\|_2^2}{\mu T^2}$. Finally, owing to the fact $-f_\mu(\mathsf{w}^*) \leq -f_\mu(\mathsf{w}_T)$, we get $f_\mu(u) - f_\mu(\mathsf{w}^*) \leq \frac{2\|u\|_2^2}{\mu T^2}$. □

The above theorem immediately implies the following result for MaxIB.

> **Theorem B.3.** *Suppose some value $\beta > 0$ is known and satisfies $\beta/c \leq r_{\mathsf{opt}} \leq \beta$ for some constant $c$. Then, letting $\mu = \frac{\varepsilon\beta}{2c\log m}$ and $T = \lceil\frac{4\sqrt{2}\alpha c\sqrt{\log m}}{\varepsilon}\rceil = O(\frac{\sqrt{\log m}\alpha}{\varepsilon})$, the point $\mathsf{w}^* = \mathtt{MaxIBConvexSolver}(f_\mu, T)$ is the center of an inscribed ball whose radius is at least $(1-\varepsilon)r_{\mathsf{opt}}$. The total running time is*
> $$O\big(md \cdot \sqrt{\log m}\alpha/\varepsilon\big) \ .$$
>
> *In addition, $\mathtt{MaxIBConvexSolver}$ is a parallel algorithm: it converges in $O\big(\sqrt{\log m}\alpha/\varepsilon\big)$ iterations, each consisting of a matrix-vector multiplicative that is highly parallelizable.*

*Proof of Theorem B.3.* By Lemma 3.3, we have $\|x^*\| \leq 2R \leq 2\alpha r_{\mathsf{opt}} \leq 2\alpha\beta$.

Our choice of $T = \lceil\frac{4\sqrt{2}\alpha c\sqrt{\log m}}{\varepsilon}\rceil = \lceil\frac{4\alpha\beta}{\sqrt{\mu\varepsilon\beta/c}}\rceil$ implies that according to Theorem B.2

$$f_\mu(x^*) - f_\mu(\mathsf{w}^*) \leq \frac{2\|x^*\|_2^2}{\mu T^2} \leq \frac{2(2\alpha\beta)^2}{\mu T^2} \leq \frac{\varepsilon\beta}{2c} \leq \frac{\varepsilon r_{\mathsf{opt}}}{2} \ .$$

On the other hand, Proposition B.1.a implies that $f_\mu(x^*) \geq f(x^*) = r_{\mathsf{opt}}$. Together, we obtain that $f_\mu(\mathsf{w}^*) \geq (1-\varepsilon/2)r_{\mathsf{opt}}$ and after applying Proposition B.1.a again we have

$$f(\mathsf{w}^*) \geq (1-\varepsilon/2)r_{\mathsf{opt}} - \mu\log m \geq (1-\varepsilon/2)r_{\mathsf{opt}} - \frac{\varepsilon\beta}{2c} \geq (1-\varepsilon)r_{\mathsf{opt}} \ . \quad \square$$

In Appendix C we will show that the preprocessing step of computing $\beta$ only requires a running time of $O(md\alpha\log\alpha)$, so applying $\mathtt{MaxIBConvexSolver}$ once after the preprocessing solves MaxIB.

## C  Preprocessing Step For MaxIB

When solving MaxIB in Section 3 and Appendix B we have assumed that a constant approximation to $r_{\mathsf{opt}}$ is already given. We prove in this section that it is easy to obtain such a result.

In general, there are two ways to get a constant approximation to $r_{\mathsf{opt}}$. The first one is to use the preprocessing procedure of [XSX06, Lemma 22] to deduce a good starting point. We restate their lemma in our language as follows:

**Lemma C.1** ([XSX06]). *In a total running time of $O(dm\alpha\log\alpha)$, one can find a point $x_0 \in P$ satisfying $f(x_0) \leq r_{\mathsf{opt}} \leq 3f(x_0)$. In other words, this is a 3-approximation to MaxIB.*

Unfortunately, this preprocessing algorithm of [XSX06] is still a geometry-based algorithm so is not much simpler than their generic one. We now propose an alternative, very simple preprocessing strategy. Informally speaking, we have

**Lemma C.2.**

1. *By continuously halving $\beta$, and applying $\mathtt{MaxIBConvexSolver}$ as an oracle, with an overhead running time $O(dm\sqrt{\log m}\alpha\log\alpha)$ one can get a constant approximation to $r_{\mathsf{opt}}$.*

2. *By continuously halving $\beta$, and applying $\mathtt{MaxIBSPSolver}$ as an oracle, with an overhead running time $\widetilde{O}(dm + \sqrt{d}m\alpha)$ one can get a constant approximation to $r_{\mathsf{opt}}$.*

The advantage of using the preprocessing as above is the ability to call our well-established optimization oracles directly, without recurring to a third-party algorithm. In the rest of this section, we only demonstrate how to achieve the first goal in Lemma C.2. The second goal is very similar.

**Algorithm 5** `MaxIBConvexSolverFull`$(A, b, \varepsilon, \alpha)$

**Input:** Bounded polyhedron $P = \{x \colon Ax + b \geq 0\}$; error constant $\varepsilon$, and aspect ratio upper bound $\alpha$.

**Output:** $\mathsf{w}^*$ such that $f(\mathsf{w}^*) \geq (1 - \varepsilon) r_{\mathsf{opt}}$.

1: $\beta \leftarrow B$ and $T_0 \leftarrow \lceil 16\alpha\sqrt{\log m} \rceil$.
2: **loop**
3:     $\mu \leftarrow \frac{\beta}{8 \log m}$.
4:     $\mathsf{w}^* \leftarrow$ `MaxIBConvexSolver`$(f_\mu, T_0)$.
5:     **if** $f_\mu(\mathsf{w}^*) \leq \frac{\beta}{4}$ **then**
6:        $\beta \leftarrow \beta/2$.
7:     **else**
8:        **break**
9:     **end if**
10: **end loop**

                                                  $\diamond$ owing to Lemma C.3, we now know $\beta/8 \leq r_{\mathsf{opt}} \leq \beta$.

11: $\mu \leftarrow \frac{\varepsilon \beta}{16 \log m}$ and $T \leftarrow \lceil \frac{32\sqrt{2}\alpha\sqrt{\log m}}{\varepsilon} \rceil$.
12: **return** `MaxIBConvexSolver`$(f_\mu, T)$.

**Formal Statement and Proof of Lemma C.2.** Recall that at the very beginning we have $r_{\mathsf{opt}} \leq B$ according to Fact 3.1. Accordingly, we now set $\beta \leftarrow B$ as an upper bound to $r_{\mathsf{opt}}$, and gradually decrease it. In each outer iteration, we run `MaxIBConvexSolver`$(f_\mu, T_0)$ for a value of $\mu = \Theta(\beta/\log m)$ and an iteration count $T_0 = \Theta(\alpha\sqrt{\log m})$. The following lemma guarantees that we either (1) find a certificate that $r_{\mathsf{opt}} \leq \beta/2$ so we can decrease $\beta$ to $\beta/2$, or (2) find a constant approximation to $r_{\mathsf{opt}}$.

**Lemma C.3.** *Suppose that $r_{\mathsf{opt}} \leq \beta$, and $\mathsf{w}^* = $ `MaxIBConvexSolver`$(f_\mu, T_0)$ for $\mu = \frac{\beta}{8 \log m}$ and $T_0 = \lceil 16\alpha\sqrt{\log m} \rceil$. Now,*
- *If $f_\mu(\mathsf{w}^*) \leq \frac{\beta}{4}$, then $r_{\mathsf{opt}} \leq \frac{\beta}{2}$; or*
- *If $f_\mu(\mathsf{w}^*) > \frac{\beta}{4}$, then $r_{\mathsf{opt}} \geq \frac{\beta}{8}$.*

*Proof.* As in the proof of Theorem B.3, we have $\|x^*\| \leq 2\alpha r_{\mathsf{opt}}$. Since $\mu = \frac{\beta}{8 \log m}$, Theorem B.2 implies that after $T_0 = \lceil 16\alpha\sqrt{\log m} \rceil$ iterations, we have

$$f_\mu(x^*) - f_\mu(\mathsf{w}^*) \leq \frac{2\|x^*\|_2^2}{\mu T_0^2} \leq \frac{8\alpha^2 r_{\mathsf{opt}}^2}{\mu T_0^2} \leq \frac{64 \log m \alpha^2 r_{\mathsf{opt}}^2}{256 \beta \log m \alpha^2} \enspace.$$

Because we are given that $r_{\mathsf{opt}} \leq \beta$, this further implies $f_\mu(x^*) - f_\mu(\mathsf{w}^*) \leq \frac{1}{4} r_{\mathsf{opt}} \leq \frac{1}{4}\beta$. We are now ready to prove the two claims respectively.

If $f_\mu(\mathsf{w}^*) \leq \frac{\beta}{4}$, then owing to Proposition B.1.a we have $r_{\mathsf{opt}} = f(x^*) \leq f_\mu(x^*) \leq f_\mu(\mathsf{w}^*) + \frac{1}{4}\beta \leq \frac{1}{2}\beta$ as desired.

If $f_\mu(\mathsf{w}^*) > \frac{\beta}{4}$, then owing to Proposition B.1.a we have $r_{\mathsf{opt}} = f(x^*) \geq f_\mu(x^*) - \mu \log m = f_\mu(x^*) - \frac{\beta}{8} \geq f_\mu(\mathsf{w}^*) - \frac{\beta}{8} \geq \frac{\beta}{8}$ as desired. □

Therefore, the following theorem is a direct consequence of the lemma above.

**Theorem C.4.** `MaxIBConvexSolverFull`$(A, b, \varepsilon)$ *produces a* $(1 − \varepsilon)$ *approximate solution to the maximum inscribed ball problem, and requires a total of* $O(\sqrt{\log m}\alpha \log \alpha + \frac{\sqrt{\log m}\alpha}{\varepsilon})$ *iterations of* `MaxIBConvexSolver`*. Since each iteration is dominated by a matrix vector multiplication, this is a total running time of*

$$O\Big(md \cdot \sqrt{\log m}\alpha \big(\log \alpha + \frac{1}{\varepsilon}\big)\Big) \ .$$

*Proof.* The last call of `MaxIBConvexSolverFull` clearly requires $T = O(\frac{\sqrt{\log m}\alpha}{\varepsilon})$ iterations. Therefore, it suffices for us to show that $\beta$ is at most halved $O(\log \alpha)$ times. Noticing that we have begun with $\beta = B$, and as we always have $\beta \geq r_{\mathsf{opt}}$, this halving process cannot happen for more than $O(\log \frac{B}{r_{\mathsf{opt}}})$ times.

Since $B$ equals to the distance between $O$ and some hyperplane —say, $H_1$— we can draw the segment between $O$ and *any* point $Q$ on the hyperplane of $H_1$ that is also contained in the polyhedron $P$.[7] This segment $\overline{OQ}$ is completely contained in the polyhedron $P$, and therefore must be of length no more than $2R$. We conclude now that $B \leq |\overline{OQ}| \leq 2R$, which implies $O(\log \frac{B}{r_{\mathsf{opt}}}) = O(\log \alpha)$. This finishes the proof on the upper bound of the number of halving steps on $\beta$, and therefore on the total running time of `MaxIBConvexSolverFull`. $\square$

## D MinEB: From Geometry to Convex Smooth Optimization

In this section we rewrite our saddle-point formulation of MaxIB in (4.1) as a convex smooth optimization problem. We define

$$f(x) \stackrel{\text{def}}{=} -\min_{y \in \mathbb{R}^d} \Big\{\frac{1}{2}\sum_i x_i \|y - a_i\|_2^2\Big\} = -\min_{y \in \mathbb{R}^d} \Big\{\frac{1}{2}\|y\|_2^2 - \langle y, \sum_i x_i a_i\rangle + \frac{1}{2}\sum_i x_i \|a_i\|_2^2\Big\}$$

$$= \frac{1}{2}\|\sum_i x_i a_i\|_2^2 - \frac{1}{2}\sum_i x_i \|a_i\|_2^2 = \frac{1}{2}\|Ax\|_2^2 - \frac{1}{2}\sum_i x_i \|a_i\|_2^2 \ . \tag{D.1}$$

Therefore, it suffices for us to minimize $f(x)$ over the simplex $x \in \Delta_n$. We now claim an important property about $f(x)$:

**Lemma D.1.** *We have* $\|\nabla f(x_1) - \nabla f(x_2)\|_\infty \leq \|x_1 - x_2\|_1$.

*Proof.* This can be done by directly computing

$$\|\nabla f(x_1) - \nabla f(x_2)\|_\infty = \|A^T A x_1 - A^T A x_2\|_\infty \leq \max_{i,j \in [n]} |(A^T A)_{i,j}| \cdot \|x_1 - x_2\|_1 \ .$$

Now, since $|(A^T A)_{i,j}| = |\langle a_i, a_j\rangle| \leq 1$, we arrive at the desired inequality. $\square$

In the optimization language, Lemma D.1 says that $f(\cdot)$ is 1-*smooth with respect to the* $\ell_1$ *norm*. Therefore, we have formalized MinEB directly into a smooth convex minimization problem over the simplex $\Delta_n$.[8]

---

[7]Here, we have assumed that the halfspaces are given with no redundancy. That is, for all halfspaces $H_j$, there must exist some point $Q$ on the hyperplane $H_j$ which is also contained in the polyhedron $P$. This assumption was also implicitly made by [XSX06] when they are searching for the starting point. If this assumption is removed, the $\log \alpha$ dependency will be replaced with $\log \frac{B}{r_{\mathsf{opt}}}$ in our theorem.

[8]It is perhaps interesting to note that some prior works, such as the theoretical result [SVZ11] and the empirical result [ZTS05], rely on the perhaps more involved non-smooth optimization techniques. We hope that our simplification provides better insight into the practical solution of this problem.

**Algorithm 6** `MinEBConvexSolver`$(f, T)$

**Input:** $f$ the objective defined in (D.1), and $T$ the number of iterations.
**Output:** $\bar{y} \in \mathbb{R}^d$.
1: Define function $V_x(y) \stackrel{\text{def}}{=} \sum_{i \in [n]} y_i \log(y_i/x_i)$.
   $\diamond$ *known as the Bregman divergence function for $\ell_1$ space [BN13]*
2: $\mathsf{v}_0 = \mathsf{w}_0 = \mathsf{z}_0 \leftarrow (\frac{1}{n}, \ldots, \frac{1}{n})$.
3: $L \leftarrow \max_i \|a_i\|_2^2$.
4: $\bar{\mathsf{v}} \leftarrow 0$.
5: **for** $k \leftarrow 0$ **to** $T - 1$ **do**
6: $\quad \alpha_{k+1} \leftarrow \frac{k+2}{2L}$, and $\tau_k \leftarrow \frac{1}{\alpha_{k+1} L} = \frac{2}{k+2}$.
7: $\quad \mathsf{v}_{k+1} \leftarrow \tau_k \mathsf{z}_k + (1 - \tau_k) \mathsf{w}_k$.
8: $\quad \mathsf{w}_{k+1} \leftarrow \arg\min_{\mathsf{w}} \left\{ \frac{L}{2} \|\mathsf{w} - \mathsf{v}_{k+1}\|_1^2 + \langle \nabla f(\mathsf{v}_{k+1}), \mathsf{w} - \mathsf{v}_{k+1} \rangle \right\}$.
   $\diamond$ *see Appendix E for implementation details*
9: $\quad \mathsf{z}_{k+1} \leftarrow \arg\min_{\mathsf{z}} \left\{ V_{\mathsf{z}_k}(\mathsf{z}) + \langle \alpha_{k+1} \nabla f(\mathsf{v}_{k+1}), \mathsf{z} - \mathsf{z}_k \rangle \right\}$. $\diamond$ *see Fact E.1 for details*
10: $\quad \bar{\mathsf{v}} \leftarrow \bar{\mathsf{v}} + \alpha_{k+1} \mathsf{v}_{k+1}$.
11: **end for**
12: $\bar{\mathsf{v}} \leftarrow \bar{\mathsf{v}} / (\sum_{k=1}^T \alpha_k)$.
13: **return** $A\bar{\mathsf{v}}$.

---

**Our Algorithm.** Using the $\ell_1$ smoothness (cf. Lemma D.1) of our objective $f(x)$, we can again apply the accelerated gradient method of Nesterov [Nes05b]. We again adopt the version from [AO14] as the template for the accelerated gradient method, and write our `MinEBConvexSolver` in Algorithm 6.

In fact, the original result of Nesterov [Nes05b] directly implies that, as long as the number of iterations $T \geq \Omega(\sqrt{\log n}/\sqrt{\varepsilon})$, the choice of $x = \mathsf{w}_T \in \Delta_n$ from `MinEBConvexSolver` gives a $(1 + \varepsilon)$-approximate minimizer of $f(x)$ —that is, it satisfies $f(x) \geq -(1 - \varepsilon)\mathsf{OPT}$. Unfortunately, this value $x = \mathsf{w}_T$ does not explicitly provide any solution to MinEB: our goal is instead to find a point $\bar{y} \in \mathbb{R}^d$ that is the center of some (approximately) minimum enclosing ball.

To solve this problem, we collect the history of the vectors $\mathsf{v}_k$ in the entire execution of `MinEBConvexSolver`, define $\bar{\mathsf{v}}$ to be its average, and output $\bar{y} = A\bar{\mathsf{v}}$. Our next theorem shows that $\bar{y}$ is an approximate solution to MinEB.

> **Theorem D.2.** `MinEBConvexSolver`$(f, T)$ *produces a* $(1 + \varepsilon)$ *approximate solution to the minimum enclosing ball problem if* $T \geq \frac{\sqrt{32 \log n}}{\sqrt{\varepsilon}}$. *This totals to* $T = O(\sqrt{\log n}/\sqrt{\varepsilon})$ *iterations with a running time of*
> $$O\left(nd \cdot \frac{\sqrt{\log n}}{\sqrt{\varepsilon}}\right) \enspace .$$
> *In addition,* `MinEBConvexSolver` *is a parallel algorithm: it converges in* $O(\sqrt{\log n}/\sqrt{\varepsilon})$ *iterations, each dominated by a matrix-vector multiplicative that is highly parallelizable.*

*Proof.* (There are multiple versions of the accelerated gradient method, and our `MaxIBConvexSolver` is written from the version provided in [AO14].)

From the proof of [AO14, Lemma 4.3] —that is, the key lemma of the classical convergence proof of the accelerated gradient method— we know that for every $u \in \Delta_n$,

$$\alpha_{k+1} \langle \nabla f(\mathsf{v}_{k+1}), \mathsf{v}_{k+1} - u \rangle$$
$$\leq (\alpha_{k+1}^2 L - \alpha_{k+1}) f(\mathsf{w}_k) - (\alpha_{k+1}^2 L) f(\mathsf{w}_{k+1}) + \alpha_{k+1} f(\mathsf{v}_{k+1}) + (V_{\mathsf{z}_k}(u) - V_{\mathsf{z}_{k+1}}(u))$$

Since our choice of $\alpha_k = \frac{k+1}{2L}$ ensures that $\alpha_k^2 L = \alpha_{k+1}^2 L - \alpha_{k+1} + \frac{1}{4L}$, we can telescope the above inequality with $k = 0, 1, \ldots, T-1$ and deduce that

$$\sum_{k=1}^{T} \alpha_k(\langle \nabla f(\mathsf{v}_k), \mathsf{v}_k - u \rangle - f(\mathsf{v}_k)) \leq -\alpha_T^2 L f(\mathsf{w}_T) - \sum_{k=1}^{T-1} \frac{1}{4L} f(\mathsf{w}_k) + \big(V_{\mathsf{z}_0}(u) - V_{\mathsf{z}_T}(u)\big) \ .$$

Using $\sum_{k=1}^{T} \alpha_k = \frac{T(T+3)}{4L} = \alpha_T^2 L + \frac{T-1}{4L}$, $f(\mathsf{w}_k) \geq -\mathsf{OPT}$, and $V_{\mathsf{z}_0}(u) \leq \log n$, we obtain

$$\sum_{k=1}^{T} \alpha_k(\langle \nabla f(\mathsf{v}_k), \mathsf{v}_k - u \rangle - f(\mathsf{v}_k)) \leq \sum_{k=1}^{T} \alpha_k \mathsf{OPT} + \log n \ . \tag{D.2}$$

Next, using the definition of $\nabla f(\cdot)$, we have

$$\langle \nabla f(\mathsf{v}_k), \mathsf{v}_k - u \rangle - f(\mathsf{v}_k)$$
$$= \left\langle A^T A \mathsf{v}_k - \frac{1}{2}(\|a_1\|_2^2, \ldots, \|a_n\|_2^2), \mathsf{v}_k - u \right\rangle - \frac{1}{2}\|A\mathsf{v}_k\|_2^2 + \frac{1}{2}\sum_i \mathsf{v}_{k,i} \|a_i\|_2^2$$
$$= \frac{1}{2}\|A\mathsf{v}_k\|_2^2 - \langle A\mathsf{v}_k, Au \rangle + \frac{1}{2}\sum_i u_i \|a_i\|_2^2 \ . \tag{D.3}$$

Let us now define $\alpha \stackrel{\text{def}}{=} \sum_{k=1}^{T} \alpha_k$, $\bar{\mathsf{v}} = \sum_{k=1}^{T} \frac{\alpha_k}{\alpha} \mathsf{v}_k$ and $\bar{y} \stackrel{\text{def}}{=} A\bar{\mathsf{v}}$, then, for every $u \in \Delta_n$,

$$\frac{1}{2}\sum_i u_i \|\bar{y} - a_i\|_2^2 = \frac{1}{2}\|A\bar{\mathsf{v}}\|_2^2 - \langle A\bar{\mathsf{v}}, Au \rangle + \frac{1}{2}\sum_i u_i \|a_i\|_2^2$$
$$\stackrel{\text{①}}{\leq} \sum_{k=1}^{T} \frac{\alpha_k}{\alpha}\left(\frac{1}{2}\|A\mathsf{v}_k\|_2^2 - \langle A\mathsf{v}_k, Au \rangle + \frac{1}{2}\sum_i u_i \|a_i\|_2^2\right)$$
$$\stackrel{\text{②}}{\leq} \mathsf{OPT} + \frac{\log n}{\alpha} < \mathsf{OPT} + \frac{4L \log n}{T^2} = \mathsf{OPT} + \frac{4 \log n}{T^2} \ .$$

Above, ① uses the convexity of $\|\cdot\|_2^2$, and ② uses (D.2) and (D.3).

Finally, using the fact that $1 \leq 8\mathsf{OPT}$ from Fact 4.1, we conclude that after $T \geq \frac{\sqrt{32 \log n}}{\sqrt{\varepsilon}}$ iterations, we have

$$\frac{1}{2}\sum_i u_i \|\bar{y} - a_i\|_2^2 \leq \mathsf{OPT} \cdot (1 + \varepsilon) = \frac{1}{2} R_{\mathsf{opt}}^2 \cdot (1 + \varepsilon) \ .$$

If we maximize the left hand side with respect to all $u \in \Delta_n$, and denote by $R$ the minimum radius of the enclosing ball centered at $\bar{y}$, we immediately have $R \leq (1 + \varepsilon) R_{\mathsf{opt}}$.

The fact that each iteration of `MinEBConvexSolver` can be implemented to run in linear $O(nd)$ time can be found in Appendix E. □

# E  Implementation Details for `MinEBConvexSolver`

In this section we explain how to efficiently implement Line 8 and 9 of `MinEBConvexSolver` to run in linear $O(nd)$ time.

This is so for Line 9 because, as shown below, the heaviest computational component of Line 9 is $A^T(A\mathsf{v}_{k+1})$, implementable by two matrix-vector multiplications.

**Fact E.1.** *Line 9 of* `MinEBConvexSolver` *can be implemented as follows.*
- $g \leftarrow A^T A \mathsf{v}_{k+1} - \frac{1}{2}(\|a_1\|_2^2, \ldots, \|a_n\|_2^2)$, and
- $\mathsf{z}_{k+1} \leftarrow \frac{\mathsf{z}_k \cdot e^{-\alpha_{k+1} g_i}}{Z}$, *where $Z > 0$ is the normalization constant that ensures $\mathbb{1}^T \mathsf{z}_{k+1} = 1$.*

The above fact can be verified by taking the derivative and showing that $\mathsf{z}_{k+1}$ is indeed the minimizer. Since this is a classical step in optimization (see for instance [AO14]), we ignore it in this version of the paper.

We next show the same for Line 8. Recall that Line 8 is written as $\mathsf{w}_{k+1} \leftarrow \arg\min_\mathsf{w} \{\frac{L}{2}\|\mathsf{w} - \mathsf{v}_{k+1}\|_1^2 + \langle \nabla f(\mathsf{v}_{k+1}), \mathsf{w} - \mathsf{v}_{k+1}\rangle\}$, where $\nabla f(\mathsf{v}_{k+1})$ can be computed in $O(nd)$ time. Therefore, it suffices to prove the following:

**Lemma E.2.** *The minimizer $\arg\min_{z\in\Delta_n}\{\frac{L}{2}\cdot\|z - v\|_1^2 + \langle \xi, z\rangle\}$ can be computed in $O(n)$ total time.*

The rest of this section is devoted to proving this above lemma. Without loss of generality, let us assume that $\xi_1 \geq \xi_2 \geq \cdots \geq \xi_n$.

We consider *states* $s$ of the form $s = (i, a) \in [n] \times [0, 1]$, where $i \in [n]$ and $a \in [0, 1]$. We say that a state $s_1 = (i_1, a_1)$ is *smaller* than $s_2 = (i_2, a_2)$ if $i_1 < i_2$ or $i_1 = i_2$ and $a_1 < a_2$. We define two functions $g(s)$ and $h(s)$ over all possible states.

$$\forall s = (i, a), \quad \text{define} \quad g(s) \stackrel{\text{def}}{=} a \cdot v_i + \sum_{j=1}^{i-1} v_j \quad \text{and} \quad h(s) \stackrel{\text{def}}{=} \frac{\xi_i - \xi_n}{4L} \enspace.$$

It is now clear that $g(s)$ is monotonically non-decreasing as $s$ increases, while $h(s)$ is monotonically non-increasing as $s$ increases. In addition, when $s = (1, 0)$, we have $h(s) \geq 0 = g(s)$; when $s = (n, 0)$ we have $g(s) \geq 0 = h(s)$.

Therefore, one can find —for instance, by binary search— a state $s^*$ between $(1, 0)$ and $(n, 0)$ satisfying that it is the *largest* state satisfying $g(s^*) \leq h(s^*)$. Denote this state $s^*$ as $(i^*, a^*)$, and let $\delta \stackrel{\text{def}}{=} g(s^*) \in [0, h(s^*)]$.

We wish to prove next that

$$z^* = \big(0, 0, \ldots, 0, (1 - a^*) \cdot v_{i^*}, v_{i^*+1}, \ldots, v_{n-1}, v_n + \delta\big)$$

is the minimizer $\arg\min_{z\in\Delta_n} \{\frac{L}{2}\cdot\|z - v\|_1^2 + \langle \xi, z\rangle\}$. Since it is easy to verify that $z^* \in \Delta_n$ by the definition of $\delta$, to show $z^*$ is the minimizer it suffices to verify the following claim

**Claim E.3.** *The zero vector is a valid subgradient of the function $\phi(z) \stackrel{\text{def}}{=} \frac{L}{2} \cdot \|z - v\|_1^2 + \langle \xi, z\rangle$ at point $z = z^*$ with constraint $\Delta_n$.*

*Proof.* We first compute that the subgradients of $\phi(z)$ with respect to constraint $\Delta_n$ consist of all vectors of the form

$$\partial\phi(z) = L \cdot \|z - v\|_1 \cdot (q_1, q_2, \ldots, q_n) + \xi + c \cdot (1, 1, \ldots, 1) - \sum_{i=1}^n t_i \mathsf{e}_i \enspace,$$

where $c$ can be any real value, and for each $i \in [n]$,
- we have $q_i = +1$ if $z_i > v_i$, $q_i = -1$ if $z_i < v_i$, or $q_i$ can be any value in $[-1, 1]$ if $z_i = v_i$;
- we have $t_i = 0$ if $z_i > 0$, or $t_i$ can be any non-negative value if $z_i = 0$.

(Above, $c$ is the Lagrangian multiplier for the constraint $\mathbb{1}^T z = 0$, and each $t_i$ is the Lagrangian multiplier for the constraint $z_i \geq 0$.)

Let us now verify that 0 is one such valid subgradient at point $z = z^*$. We define

- $c = -(2L\delta + \xi_n)$.
- $q_i = -1$ for all $i \leq i^*$.
- $q_n = +1$.
- $q_i = -\frac{c+\xi_i}{2L\delta}$ for all $i \in \{i^*+1, i^*+2, \ldots, n-1\}$.
- $t_i = 0$ for all $i > i^*$.
- $t_i = \xi_i + c + 2L\delta q_i$ for all $i \leq i^*$.

Recall that $\|z^* - v\|_1 = \delta + a^* v_{i^*} + \sum_{j=0}^{i^*-1} v_i = \delta + g(s^*) = 2\delta$ by the definition of $z^*$ and $\delta$.

For each $i \leq i^*$, we have $L \cdot \|z^* - v\|_1 \cdot q_i + \xi_i + c - t_i = 0$ by the definition of $t_i$. For each $i \in \{i^*, i^*+1, \ldots, n-1\}$, we have $L \cdot \|z^* - v\|_1 \cdot q_i + \xi_i + c - t_i = 0$ by the definition of $q_i$. For $i = n$, we have $L \cdot \|z^* - v\|_1 \cdot q_i + \xi_i + c - t_i = 2L\delta + \xi_n - (2\delta L - \xi_n) = 0$. Therefore, we have verified that

$$L \cdot \|z^* - v\|_1 \cdot (q_1, q_2, \ldots, q_n) + \xi + c \cdot (1, 1, \ldots, 1) - \sum_{i=1}^{n} t_i \mathbf{e}_i = 0 \ .$$

Next, for each $i \in \{i^* + 1, i^* + 2, \ldots, n-1\}$, we want to show $q_i \in [-1, 1]$, satisfying the requirement on $q$. To begin with, we have $q_i = -\frac{-(2L\delta+\xi_n)+\xi_i}{2L\delta} = 1 - \frac{\xi_i - \xi_n}{2L\delta} \leq 1$. To show $q_i \geq -1$ as well, we assume by way of contradiction that $q_i < -1$. This implies $1 - \frac{\xi_i - \xi_n}{2L\delta} < -1$ which gives $\delta < \frac{\xi_i - \xi_n}{4L}$. Since $\xi_{i*} \geq \xi_i$, we immediately have

$$g(s^*) = \delta < \frac{\xi_i - \xi_n}{4L} \leq \frac{\xi_{i^*} - \xi_n}{4L} = h(s^*) \ .$$

There are now two cases. Recall that the state $s^* = (i^*, a^*)$.
- If $a^* < 1$, we can consider $s = (i^*, a^* + \epsilon)$. When $\epsilon > 0$ is sufficiently small, we must also have $g(s) < h(s)$, contradicting to the definition that $s^*$ is the maximal state satisfying $g(s) \leq h(s)$.
- If $a^* = 1$, we can consider state $s = (i^* + 1, 0)$. It is easy to see that $g(s) = g(s^*)$, but the above inequality tells us $g(s) < \frac{\xi_i - \xi_n}{4L} \leq \frac{\xi_{i^*+1} - \xi_n}{4L} = h(s)$, again contradicting to the definition of $s^*$.

In sum, we must have $q_i \in [-1, 1]$.

It is now only left to show that $t_i \geq 0$ for all $i \leq i^*$. Indeed, we can compute that $t_i = \xi_i + c + 2L\delta q_i = \xi_i - (2L\delta + \xi_n) - 2L\delta = \xi_i - \xi_n - 4L\delta = 4L(\frac{\xi_i - \xi_n}{4L} - \delta) \geq 0$. Here, the last inequality is due to $g(s^*) = \delta \leq h(s^*) = \frac{\xi_{i^*} - \xi_n}{4L}$. □

Finally, we have finished proving that $z^*$ is the desired minimizer.

It is now easy to see that the implementation of the above binary search procedure can be implemented to run in $O(n)$ total time plus the time needed for sorting. This totals to $O(n \log n)$ so only yields a weaker version of Lemma E.2.

Although this is efficient enough for practical applications, one can improve this total running time to $O(n)$ by using the $O(n)$ median-finding algorithm to avoid sorting, using similar ideas from [SVZ11, Appendix B]. We leave it an exercise for interested readers to figure out the details. □